\documentclass[12pt]{iopart}

\usepackage{iopams} 
\expandafter\let\csname equation*\endcsname\relax
\expandafter\let\csname endequation*\endcsname\relax
\usepackage{amsmath}
\usepackage{amsthm}
\usepackage{graphicx}
\usepackage[pdfpagelabels=false,colorlinks=true,allcolors=black]{hyperref}

\newcommand{\eps}{\varepsilon}
\newcommand{\R}{\mathbb{R}}
\newcommand{\Z}{\mathbb{Z}}
\renewcommand{\vec}{\mathbf}
\renewcommand{\mid}{\,|\,}
\newcommand{\inner}[2]{\left\langle #1 \,|\, #2 \right\rangle}
\newcommand{\bra}[1]{\left\langle #1 \right|}
\newcommand{\ket}[1]{\left| #1 \right\rangle}
\newcommand{\Exp}[1]{\mathbb{E}\!\left[ #1 \right]}
\newcommand{\Psei}{\widetilde{\mathbb{E}}}
\newcommand{\Pse}[1]{\widetilde{\mathbb{E}}\!\left[ #1 \right]}
\newcommand{\Pnull}{P_0}
\newcommand{\Pplanted}{P_1}

\newcommand{\ignore}[1]{\relax}

\newcommand{\E}{\mathbb{E}}

\newtheorem{theorem}{Theorem}

\begin{document}

\title[Disordered Systems Insights on Computational Hardness]{Disordered Systems Insights on Computational Hardness}

\author{David Gamarnik$^1$, Cristopher Moore$^2$, Lenka Zdeborov\'a$^3$}

\address{$^1$ Operations Research Center and Sloan School of Management, MIT, Cambridge, MA 02139}
\address{$^2$ Santa Fe Institute, Santa Fe, NM 87501, USA}
\address{$^3$ SPOC Laboratory, École Polytechnique Fédérale de Lausanne (EPFL), Route Cantonale, CH-1015 Lausanne, Switzerland}
\ead{$^1$gamarnik@mit.edu, $^2$moore@santafe.edu, $^3$lenka.zdeborova@epfl.ch}
\vspace{10pt}

\begin{abstract}
In this review article we discuss connections between the physics of
disordered systems, phase transitions in inference problems, and computational
hardness. We introduce two models representing the behavior of glassy
systems, the spiked tensor model and the generalized linear model. We discuss the random (non-planted) versions of these problems as prototypical optimization problems, as well as the planted versions (with a hidden solution) as prototypical problems in statistical inference and learning. Based on ideas from physics, many of these problems have transitions where they are believed to jump from easy (solvable in polynomial time) to hard (requiring exponential time). We  discuss several emerging ideas in theoretical computer science and statistics that provide rigorous evidence for hardness by proving that large classes of algorithms fail in the conjectured hard regime. This includes the overlap gap property, a particular mathematization of clustering or dynamical symmetry-breaking, which can be used to show that many algorithms that are local or robust to changes in their input fail. We also discuss the  sum-of-squares hierarchy, which places bounds on proofs or algorithms that use low-degree polynomials such as standard spectral methods and semidefinite relaxations, including the Sherrington-Kirkpatrick model. Throughout the manuscript we present connections to the physics of disordered systems and associated replica symmetry
breaking properties. 
\end{abstract}

%
%
%
%
%

\section{Introduction}\label{sec:intro}

Computational complexity theory~\cite{nature_of_computation} aims to answer the question of what problems can be solved by computers. More specifically, it aims to classify computational problems according to the resources (usually time or memory) needed to solve them, and how these resources scale with the problem size. Computationally hard problems are those that can be solved in principle but require prohibitively large amounts of resources, such as a running time that grows exponentially with the problem size. 

The most iconic result of computational complexity theory is the existence of so-called NP-complete problems~\cite{cook1971complexity}. These problems, of which hundreds have been identified, are all hard unless $\mathrm{P}=\mathrm{NP}$, in which case they are all easy. But if $\mathrm{P}=\mathrm{NP}$, anything which is easy to check would be easy to find. All modern cryptosystems would be breakable; it would be easy to find short proofs of unsolved mathematics problems or elegant theories to explain empirical data, without any need for insight or intuition. Even evolution would gain shortcuts: it would be easy to design proteins with certain structures, rather than having to search for them by exploring a vast space of possible amino acid sequences. This would violate many of our deepest beliefs about the nature of search, proof, and even creativity. For these and other reasons, resolving the $\mathrm{P} \ne \mathrm{NP}$ conjecture is considered the most important problem of theoretical computer science, and one of the most important open problems in mathematics more generally.

Since we believe some problems are computationally hard, the question becomes the nature of this hardness. What is it about a problem's structure that defeats  polynomial-time algorithms? 
Since the late 1980s and early 1990s (e.g.,~\cite{fu1986application,cheeseman1991really,monasson1999determining}), some researchers have looked to the physics of disordered systems as one source of hardness. This comes very naturally since, for many canonical models such as spin glasses, finding a ground state is easily shown to be NP-hard (i.e., at least as hard as any NP-complete problem). 

Physical dynamics is itself computationally limited by the locality of interactions, and physics-based algorithms such as Markov Chain Monte Carlo and simulated annealing are subject to the same limits. In glassy systems these algorithms often get stuck in metastable states, or take exponential time to cross free energy barriers. Unless there is some miraculous algorithmic shortcut for exploring glassy landscapes---which seems unlikely, except for a few isolated cases---it seems likely that no polynomial-time algorithms for these problems exist.


In this paper we review some current areas of research on the connections between theory of disordered systems and computational hardness, and attempts to make this physical intuition mathematically rigorous. We will discuss two types of computational problems: \emph{optimization problems} where one aims to minimize an objective function (such as the energy) over a set of variables, and \emph{signal recovery} or \emph{inference problems} where a signal is observed but obscured by noise, and the task is to reconstruct it (at least approximately) from these observations.  
In Section~\ref{sec:problem_defs} we define canonical examples of both these problems, stressing their relationship to
disordered systems studied in physics as well as their broad applicability to modelling various computational tasks. In Section~\ref{sec:OGP} we discuss recent results on computational hardness of
optimization problems based on the \emph{overlap gap property}, which formalizes the idea that solutions are widely separated from each other by energy barriers. Section~\ref{sec:tradeoffs} switches to signal recovery/inference problems and
presents a rather generic picture that emerges from the study of phase
transition in those problems. Finally, Section~\ref{sec:sos} discusses the 
\emph{sum-of-squares hierarchy}, another approach to proving computational lower bounds.

\section{Two problems in optimization and inference: Definitions}
\label{sec:problem_defs}


\subsection{The spiked tensor model and spin glasses}
\label{sec:spiked-tensor}

One of the models we will consider from the statistics and computational perspective is a natural variant of the spin glass model
with a ``planted signal'' to be learned or  reconstructed---physically, a low-energy state built into the landscape. It is called the \emph{spiked tensor model} or \emph{tensor PCA}, and is defined as follows. Given a hidden vector $u\in \R^N$, we observe the following tensor:
\begin{equation}
\label{eq:tensor-pca-def}
Y=\lambda u^{\otimes p}+J \, .
\end{equation}
Here $u^{\otimes p}$
is the $p$-fold tensor outer product of $u$, and $J$ is a $N \times \cdots \times N$ tensor describing the noise. We will assume that the entries $J_{i_1,\ldots,i_p}$ with $1\le i_1<i_2<\cdots<i_p\le N$ are drawn i.i.d.\ from some common distribution with mean zero and variance $\sigma^2$, such as the normal distribution $\mathcal{N}(0,1)$. The other entries of $J$ are fixed by a symmetry assumption, 
$J_{i_{\sigma(1)},\ldots,i_{\sigma(p)}}=J_{i_1,\ldots,i_p}$ for all permutations $\sigma$ of $[p]=\{1,2,\ldots,p\}$. 

We can think of  $\lambda$ as a signal-to-noise ratio, parametrizing how strongly the signal $u$ affects the observation $Y$ compared to the noise $J$. In order to look for phase transitions in the hardness of reconstructing the planted vector $u$, we will allow $\lambda$ to scale in various ways with $N$. We can also let $J$'s variance $\sigma^2$ vary with $N$, but in most of the paper we will take it to be $1$.

We can consider variants of this problem where different types of restrictions are placed on $u$. One is to take $u \in S_N$ where $S_N$ is the $N$-dimensional sphere $\{ u : \| u \|^2 = N \}$. Another choice is to take Boolean values on the $N$-dimensional hypercube or equivalently Ising spins, $u \in B_N$ where $B_N = \{\pm 1\}^N$. We can also impose sparsity by demanding that a fraction $\rho$ of $u$'s entries are nonzero, writing $u\in B_{N,\rho}$ where $B_{N,\rho}=\{u \in \{\pm 1,0\}^N: \|u\|_1=N\rho\}$. In terms of Bayesian inference, we take the uniform measure on each of these sets to be a prior on $u$.

The variant $p=2$, i.e., the spiked matrix model, is particularly widely studied. It is also known as the spiked covariance model, or as low-rank matrix estimation, since $u \otimes u$ is a rank-1 approximation of $Y$~\cite{donoho2018optimal,lesieur2017constrained}.  

The general questions to be addressed in this model are (a) can we learn, or reconstruct, the planted vector $u$ from the observation $Y$? and (b) can we do this with an efficient algorithm, i.e., one whose running time is polynomial in $N$? (We assume $p$ is a constant, so polynomial in $N$ is equivalent to polynomial in the size $N^p$ of the observed data.) Since reconstructing $u$ exactly is often impossible, we are interested in approximate reconstruction, i.e., producing an algorithmic estimate $\hat{u}=\hat{u}(Y)$ which has a nontrivial correlation with the ground truth $u$: for instance, by having an overlap $(1/N) \langle \hat{u}, u \rangle$ bounded above zero with high probability. 

Question (a) is an information-theoretic or statistical question, unconcerned with computational resources. Using the theory of Bayesian inference we can write the posterior distribution, 
\begin{align}\label{eq:Bayes-optimal}
     P(z|Y) &= \frac{1}{\mathcal {Z}} \,P(z) \,P(Y|z)  \quad {\rm where} \quad \\
P(Y|z) &= \prod_{1\le i_1<i_2<\cdots<i_p\le N} {\cal N}(Y_{i_1,\ldots,i_p} - \lambda z_{i_1}z_{i_2}\cdots z_{i_p}, 1) \, ,
\label{eq:gaussian-case}
\end{align}
where for concreteness we considered the elements of the noise $J$ to be Gaussian with variance $1$. (Due to universality properties, e.g. \cite{lesieur2017constrained}, this is not very restrictive for what follows.) 
Note that the partition function or normalization factor $\mathcal{Z}$ depends both on the observed tensor $Y$, the prior $P(z)$, and the parameters $\lambda, \sigma$ of the likelihood $P(Y|z)$. In our notation we drop this explicit dependence.  

The posterior distribution $P(z|Y)$ is an exponentially complicated object. However, for several natural loss functions including the overlap $\langle \hat{u},u \rangle$ and the $\ell_2$ error $\|\hat{u}-u\|^2$, the best possible estimator $\hat{u}$ depends only on the marginals $P(z_i|Y)$. 
Thus question (b) boils down to whether, given $Y$, we can approximate these marginals with a polynomial-time algorithm.


Another common approach in statistics is the maximum likelihood estimator\footnote{It should be noted that while the MLE and similar extremization-based approaches are very popular in statistics, they are typically suboptimal in high-dimensional settings: that is, they do not optimize the overlap or minimize the $\ell_2$ error.} (MLE) where we set $\hat{u}$ to the $z$ that maximizes $P(Y|z)$. In the Gaussian case~\eqref{eq:gaussian-case}, we have
\begin{align}
    P(Y|z) &\propto 
    \exp \left[ 
    - \frac{1}{2} \sum_{1\le i_1<i_2<\cdots<i_p\le N} \left( 
    Y_{i_1,\ldots,i_p} - \lambda z_{i_1}z_{i_2}\cdots z_{i_p}
    \right)^2 
    \right] \nonumber \\
    &= \exp \left[
    - \frac{1}{p!} \left( \frac{1}{2} \|Y\|^2 + \frac{\lambda^2}{2} \|z\|^{2p} 
    - 2 \left\langle Y, z^{\otimes p} \right\rangle \right) \right] \, ,
\end{align}
where in the limit of large $N$ we ignore terms with repeated indices, and where
\begin{align}\label{eq:spiked-tensor}
 \langle Y,z^{\otimes p}\rangle =\sum_{1\le i_1<i_2<\cdots<i_p\le N}Y_{i_1,\ldots,i_p}z_{i_1}z_{i_2}\cdots z_{i_p} \, .
\end{align}
Since $\|Y\|^2$ is fixed by the observed data, and since $\|z\|^2=N$ if $z \in S_N$ or $B_N$ (or $\rho N$ if it is in $B_{N,\rho}$) then the MLE is the $z$ that maximizes~\eqref{eq:spiked-tensor}. But this is exactly the ground state of a $p$-spin model with coupling tensor $Y$, with spherical or Ising spins if $z$ is in $S_N$ or $B_N$ respectively.


In particular, if $\lambda=0$ so that $Y=J$, we have a $p$-spin model with Gaussian random couplings and Hamiltonian
\begin{align}\label{eq:p-spin-ground-state}
E(z) = -\sum_{1\le i_1<i_2<\cdots<i_p\le N}J_{i_1,\ldots,i_p}z_{i_1}z_{i_2}\cdots z_{i_p} \, .
\end{align}
Studying the optimization landscape of this un-planted problem may seem irrelevant to the inference problem of reconstructing $u$ from $Y$. But in addition to being physically natural, as a generalization of the Sherrington-Kirkpatrick model~\cite{sherrington1975solvable} which corresponds to the case $p=2$ and $z \in B_N$, it serves both as a starting point for the inference problem and as a null model where there is no signal at all. 

Thus in addition to the \emph{reconstruction problem} where we assume that $Y$ is drawn from the planted model~\eqref{eq:tensor-pca-def} and we want to learn $u$, we will also consider the \emph{detection}  problem. That is, given $Y$, we want to determine whether it is drawn from the planted model, or the un-planted model where $Y=J$. Like reconstruction, this hypothesis testing problem may or may not be information-theoretically possible. If it is, it may or may not have a polynomial-time algorithm that succeeds with high probability.

In the literature there are many variants of the spiked tensor model. The signal can be of higher rank, i.e., $\sum_j u_j^{\otimes p}$ for multiple planted vectors $u_j$, or one can plant a subspace rather than a vector. In addition to being non-Gaussian, the noise can be nonadditive, binary or sparse. And the observation could consist of multiple tensors with different $p$ rather than a single $Y$. All these variants have their
own interest and applications; see examples in
e.g.~\cite{babacan2012sparse,lesieur2017constrained}. In what follows we will also sometimes refer to sparse versions of the spiked matrix model, such as the stochastic block model which is popular in network science as a model of community structure (see e.g.~\cite{moore_eatcs}).

\subsection{The generalized linear model and perceptrons}

Another class of problems we will consider in this paper is the generalized linear model (GLM). Again, a planted vector $u \in \R^N$ is observed through a set of noisy observations, but this time through approximate linear combinations $Y_1,\ldots,Y_P$:
\begin{align}
    Y_i \sim P_{\mathrm{out}} \!\left(Y_i \mid \sum_{a=1}^N J_{ia} u_a \right) \, . 
    \label{eq:GLM}
\end{align}
Here $J \in \R^{P \times N}$ is a known matrix whose entries are i.i.d.\ with zero mean and variance $\sigma^2$, and $P_{\mathrm{out}}$ is some noisy channel. In other words, $f(j)=\langle j,u \rangle$ is an unknown linear function from $\R^N$ to $\R$, and our goal is to learn this function---that is, to reconstruct $u$---from noisy observations of its values $f(j_1), \ldots, f(j_P)$ at $P$ random vectors where $j_i$ is the $i$th row of $J$. In machine learning we would say that the set of tuples $(j_i,Y_i)$ are the training data, and by learning $u$ we can generalize to $f(j)$ for new values of $j$. 

The main questions for the GLM are the same as for the spiked tensor model: (a) whether it is information-theoretically possible to learn the signal $u$ given $J$ and $Y$, and (b) whether there are efficient algorithms that do that. Again Bayesian inference aims at computing the marginals of a posterior 
\begin{align}
    P(z|Y,J) = \frac{1}{{\cal Z}} \,P(z) \prod_{i=1}^P P_{\mathrm{out}}\!\left( Y_i \mid \sum_{a=1}^N J_{ia} u_a \right) \, . 
 \end{align}
Here the partition function ${\cal Z}$ depends implicitly on the matrices  $Y$ and $J$ as well as on the parameters of the probability $P_{\mathrm{out}}$ and of the prior $P(z)$. As in tensor PCA, $u$ can be restricted to $S_N$, $B_N$ or some other set, and we will assume that its Bayesian prior is uniform over this set. 

Another family of estimators minimize some loss function $\ell$, perhaps with a regularization term with strength $\lambda$:
\begin{align}
    \mathcal{L}(z) = \sum_{i=1}^P \ell\!\left( Y_i, \sum_{a=1}^N J_{ia} z_a \right) 
    + \lambda \sum_{a=1}^N r(z_a) \, .
    \label{eq:ERM}
\end{align}
In a linear regression context, $J$ is the observed data and $Y$ the observed dependent variable, and~\eqref{eq:ERM} seeks to minimize the empirical risk $\ell$. A typical regularization term might be $r(z_a)=|z_a|$, giving the ``lasso'' or $L_1$ regularization $\lambda \|z\|_1$ which pushes $z$ towards sparse vectors. 

The GLM captures many versions of high-dimensional linear regression, and covers a broad range of applications and situations. In signal processing or imaging $u$ would be the $N$-dimensional signal/image to be reconstructed from measurements $Y$, where $J$ is the measurement matrix and the channel $P_{\mathrm{out}}$ typically consists of additive Gaussian noise. In compressed sensing we consider the under-determined case $N>P$, but with a sparse prior on the signal $u$. 

Just as for the spiked tensor model the signal $u$ can be seen as a planted solution to recover from $Y$ and $J$. The version of the model where the distribution of $Y$ is independent of $u$ is well known in the statistical physics literature as the perceptron. The variant with $z \in S_N$ is the spherical perceptron~\cite{gardner1988optimal}, and $z \in B_N$ gives the binary perceptron~\cite{gardner1988optimal,krauth1989storage}. The perceptron model is particularly important as its study started the line of work applying physics of disordered systems to  understanding supervised learning in artificial neural networks. The recent major success of methods based on deep learning~\cite{lecun2015deep} only added importance and urgency to this endeavour.

\section{Hardness of optimizing $p$-spin models: the overlap gap property and implications}\label{sec:OGP}

%

In this section we discuss the algorithmic hardness of  the problem (\ref{eq:p-spin-ground-state}) of 
finding near ground states of $p$-spin models using the overlap gap property (OGP). 
The OGP is a property of solution space geometry which roughly speaking says that near optimal solutions should be either close or far
from each other. It is intimately related to the replica symmetry breaking (RSB) property and the  clustering (also sometimes called shattering) property exhibited by some
constraint satisfaction problems. In fact it emerged directly as way to establish the presence of  the shattering property 
in constraint satisfaction problems~\cite{achlioptas2006solution,mezard2005clustering}.  
There are important distinctions, however, 
between RSB, clustering and OGP, which we will 
discuss as well. A survey of OGP-based methods is in~\cite{gamarnik2021overlap}. 
Our main focus is to illustrate how OGP presents a barrier to a certain class of algorithms as potential contenders for 
finding near ground states. Loosely speaking, it is the class of algorithms exhibiting input stability (noise insensitivity), thus revealing
deep and intriguing connections with a rich field of Fourier analysis of Boolean functions~\cite{o2014analysis}. Many important algorithms
are special cases of this class, including Approximate Message Passing (AMP)~\cite{gamarnik2021overlapAukosh}, 
Low-Degree Polynomials~\cite{gamarnik2020lowFOCS,wein2020optimal},
Langevin Dynamics~\cite{gamarnik2020lowFOCS}, and low-depth Boolean circuits~\cite{gamarnik2020hardness}. 
OGP was also established to be a barrier for certain types of quantum algorithms, specifically Quantum Approximate Optimization Algorithms
(QAOA)~\cite{farhi2020quantumRandom,chou2021limitations,basso2022performance}, using a slightly different implementation of the stability argument.  
We will therefore conclude that the values produced by these algorithms are bounded away from optimality. We will discuss various
extensions of the OGP, including the multi-overlap gap property (m-OGP), which will allow us to bring the algorithmic barriers to the known
algorithmic thresholds. In the case of the $p$-spin models these thresholds are achieved by AMP. 
It is entirely possible that models in the OGP regime do not admit any polynomial time algorithms, which at this stage is evidenced
by just the lack of those. Proving this say modulo $P\ne NP$ assumption does not yet appear to be within the reach of the known techniques.

\subsection{$p$-spin model,  ground states and algorithms}

We recall that our focus is the optimization problem~(\ref{eq:p-spin-ground-state}).
The optimization is over choice of $z$ in some
space $\Theta_N$ which for the purposes of this section is either $S_N$ or $B_N$. The former is referred to as spherical $p$-spin
model and the latter is called the Ising $p$-spin model.
We assume that the variance $\sigma_N^2$ of the i.i.d.\ entries of the tensor $J$ is $N^{-{(p+1)}}$. 
A series of groundbreaking works by Parisi~\cite{parisi1980sequence,MezardParisiVirasoro},  followed by 
Guerra-Toninelli~\cite{GuerraTon}, Talagrand~\cite{talagrand2006parisi}, and
Panchenko~\cite{panchenko2013parisi,panchenko2013sherrington,crisanti1992sphericalp}
led to proof of the existence and a method for computing a deterministic  limit of (\ref{eq:p-spin-ground-state})  in probability 
as $N\to\infty$. We denote this limit by  $\eta_{\rm p,OPT}$ in either case, where the choice of $\Theta_N$ will be clear from the context. 
The value of this limit arises as a solution of a certain variational 
problem over the space of one-dimensional probability measures. 
The measure which provides the solution to  this variational problem is called the Parisi measure which we denote by $\mu$.

The algorithmic goal under consideration is the goal
of constructing a solution $z\in\Theta_N$ which achieves near optimality, namely the value close to $\eta_{\rm p,OPT}$ when the tensor $J$ is given as an input. 
Ideally, we want an algorithm $\mathcal{A}$ which for every constant $\epsilon>0$ produces a solution $\hat z\triangleq \mathcal{A}(J)$
satisfying $\langle J, \hat z^{\otimes p}\rangle \ge (1-\epsilon)\eta_{\rm OPT}$ in polynomial (in $N$) time. 
This was achieved in a series
of important recent developments~\cite{subag2021following,montanari2021optimization,el2021optimization}, 
when the associated Parisi measure $\mu$ is strictly increasing. This monotonicity property is related to the OGP as we will discuss below. 

\subsection{OGP and its variants}
The following result states the presence of the OGP for the  $p$-spin  models.

\begin{theorem}\label{theorem:OGP}
For every even $p\ge 4$, $\Theta_N=B_N$ or $\Theta_N=S_N$, 
there exists $\eta_{\rm p,OGP}<\eta_{\rm p,OPT}$,  $0<\nu_1<\nu_2<1$ and $c>0$ such that
with probability at least $1-\exp(-cN)$ for large enough $N$ the following holds.  For every $z_1,z_2\in \Theta_N$ satisfying 
$\langle J,z_j^{\otimes p} \rangle \ge \eta_{\rm p,OGP}, j=1,2$ 
\begin{align*}
{1\over N}| \langle z_1,z_2\rangle | \notin (\nu_1,\nu_2).
\end{align*}
\end{theorem}
\noindent Here $\langle x,y\rangle$ denotes the inner product $\sum_{1\le i\le N}x_i y_i$. 
Namely, modulo an exponentially in $N$ unlikely event,
the normalized angle (overlap) between any two solutions with value at least $\eta_{\rm p, OGP}$ cannot fall into the interval $(\nu_1,\nu_2)$.
The model exhibits an overlap gap.

The values $\eta_{\rm p,OGP}$ and $\nu_j$ (and in fact the optimal values 
$\eta_{\rm p,OPT}$ themselves) are in general different  
for Ising and spherical models and their precise values are of no algorithmic significance. 
While the result is only known to hold for even $p\ge 4$, it is expected to hold for all $p\ge 3$. 
It is conjectured not to hold when $p=2$
\cite{MezardParisiVirasoro} for the Ising case and the AMP algorithm achieving the near ground state value in this case is effective modulo this conjecture~\cite{montanari2021optimization}. It does not hold
when $p=2$ for the spherical case for a trivial reason as in this case the problem corresponds to optimizing a quadratic form over sphere $S_N$.  
The proof of this Theorem~\ref{theorem:OGP} for the Ising case can be found in~\cite{chen2019suboptimality}, and
is obtained by a detailed analysis of the variational problem associated with pairs of   solutions $z_1,z_2$ within a certain proximity
to optimality. The proof
for the spherical case can be found in~\cite{auffinger2018energy}. 

In order to use this result as an algorithmic barrier, we need to extend this theorem to the following 
\emph{ensemble} variant of the OGP which we dub e-OGP.  For this purpose it will be convenient to assume that the distribution
of the entries of $J$ is Gaussian.
Consider an independent  pair of tensors $J,\tilde J\in \R^{N\otimes p}$ with Gaussian entries. Introduce the following  natural interpolation between the two:
$J(t)=\sqrt{1-t}J+\sqrt{t}\tilde J, t\in [0,1]$. The distribution of $J(t)$ is then identical to one of $J$ and $\tilde J$ for every $t$.

\begin{theorem}\label{theorem:e-OGP}
For every even $p\ge 4$, $\Theta_N=B_N$ or $\Theta_N=S_N$,  
for the same choice of parameters $\eta_{\rm p,OGP}, \nu_1,\nu_2$ as in Theorem~\ref{theorem:OGP}
the following holds with probability at least $1-\exp(-cN)$ for 
some $c$ and large enough $N$. For
every $t_1,t_2\in [0,1]$
and every $z_1,z_2\in \Theta_N$ satisfying $\langle J(t_j),z_j^{\otimes p}\rangle \ge \eta_{\rm p,OGP}, j=1,2$ we have 
\begin{align*}
{1\over N}| \langle z_1,z_2\rangle | \notin (\nu_1,\nu_2).
\end{align*}
Furthermore, when $t_1=0,t_2=1$, it holds ${1\over N}| \langle z_1,z_2\rangle | \in [0,\nu_1]$.
\end{theorem}
The probability event above is with respect to the joint randomness of $J$ and $\tilde J$.
Theorem~\ref{theorem:e-OGP} says that the OGP holds for pairs  of solutions with values above $\eta_{\rm p, OGP}$ \emph{across
the entire} interpolated sequence of instances $J(t)$. Furthermore, at the extremes, that is for the pair of instances $J$ and $\tilde J$,
these solutions must have overlap at most $\nu_1$. We note that the overlap value $1$ is trivially achievable when $t_1=t_2$ by taking
two identical solutions $z_1=z_2$ with value at least $\eta_{\rm p, OGP}$.
The proof for the Ising case can be found in~\cite{gamarnik2021overlapAukosh}, 
and for the spherical case in~\cite{gamarnik2020lowFOCS}, and it is a rather straightforward extension of Theorem~\ref{theorem:OGP}
by appealing to the chaos property exhibited by many glassy models~\cite{chatterjee2009disorder,chen2018disorder}.

\subsection{e-OGP as an algorithmic barrier to stable algorithms}
We now discuss how the presence of the e-OGP presents an algorithmic barrier to a class of algorithms we loosely define as \textit {stable} (noise-insensitive)
algorithms. This part will be discussed rather informally, as each concrete instantiation of the arguments is model and algorithm dependent.
We think of algorithms as mappings of the form $\mathcal{A}(J)\to\Theta_N$ which map instances (tensors) $J$ into a solution $z=\mathcal{A}(J)$ in 
the solution space $\Theta_N$. In some cases the algorithms can take advantage of an additional randomization with functions now taking the
form $\mathcal{A}(J,\omega)$, where $\omega$ is a sample corresponding to the randomization seed. For simplicity, we stick with non-randomized versions
$\mathcal{A}:\R^{N\otimes p}\to\Theta_N$. Informally, we say that the algorithm $\mathcal{A}$ is stable (noise-insensitive), if a small change in $J$ results
in a small change in the output. Namely, $\|\mathcal{A}(J_1)-\mathcal{A}(J_2)\|$ is likely to be small with respect to the natural metric on $\Theta_N$
when $\|J_1-J_2\|_2$ is small. The choice of metric on $\Theta_N$ is driven by the space itself and can be Hamming distance when $\Theta_N=B_N$
or $\mathbb{L}_2$ norm when it is $S_N$. The ``likely" is in reference to the randomness of the tensor $J$. 
The following theorem stated informally shows why the presence of the e-OGP presents a barrier to stable algorithms. 

\begin{theorem}[Informal]\label{theorem:OGP-barrier-stability}
For every stable algorithm $\mathcal{A}$ and every $\epsilon>0$, $\langle J,(\mathcal{A}(J))^{\otimes p}\rangle \le \eta_{\rm p, OGP}+\epsilon$ w.h.p. as $N\to\infty$.
\end{theorem}
Namely, this theorem states that stable algorithm cannot overcome the OGP barrier.

\begin{proof}[Proof sketch:]
We provide an outline of a simple proof of this theorem. The stability of the algorithm can sometimes be used to establish the concentration
of its value around expectation, namely that $\langle J,(\mathcal{A}(J))^{\otimes p}\rangle\approx \E{\langle J,(\mathcal{A}(J))^{\otimes p}\rangle}$ as $N\to\infty$.
This is not the case universally,  but for simplicity let's assume this for now. Then
it suffices to establish the claim $\E{\langle J,(\mathcal{A}(J))^{\otimes p}\rangle} \le \eta_{\rm p, OGP}+\epsilon$. Suppose not. Then
we have $\E{\langle J,(\mathcal{A}(J))^{\otimes p}\rangle} \ge \eta_{\rm p, OGP}+\epsilon$ implying 
$\E{\langle J(t),\mathcal{A}(J(t))\rangle} \ge \eta_{\rm p, OGP}+\epsilon$ for every $t$ in the interpolation path. We will obtain a contradiction.

By the second part of Theorem~\ref{theorem:e-OGP} we then must  have w.h.p. and in expectation
\begin{align*}
{1\over N}|\langle \mathcal{A}(J(0)),\mathcal{A}(J(1))\rangle | \le \nu_1,
\end{align*}
namely
\begin{align*}
    {1\over N}\|\mathcal{A}(J(0))-\mathcal{A}(J(1))\|_2\ge \sqrt{2-2\nu_1}.
\end{align*}
Here we assume that we use $\mathbb{L}_2$ for $\Theta_N$ and the norm of every solution produced by the algorithm is $\sqrt{N}$ (which is the case when say $\Theta_N=B_N$).
On the other hand trivially ${1\over N}|\langle \mathcal{A}(J(0)),\mathcal{A}(J(0))\rangle | =1>\nu_2$,  implying 
\begin{align*}
    {1\over N}\|\mathcal{A}(J(0))-\mathcal{A}(J(1))\|_2=0 \le \sqrt{2-2\nu_2}.
\end{align*}
Stability of the algorithm $\mathcal{A}$
implies then the existence of time $\tau$ such that 
\begin{align*}
{1\over N}|\langle \mathcal{A}(J(0)),\mathcal{A}(J(\tau))\rangle | \in (\nu_1,\nu_2),
\end{align*}
which is a contradiction to the first part of Theorem~\ref{theorem:e-OGP}. 
\end{proof}

\begin{figure}
\begin{center}
\includegraphics[width=100mm]{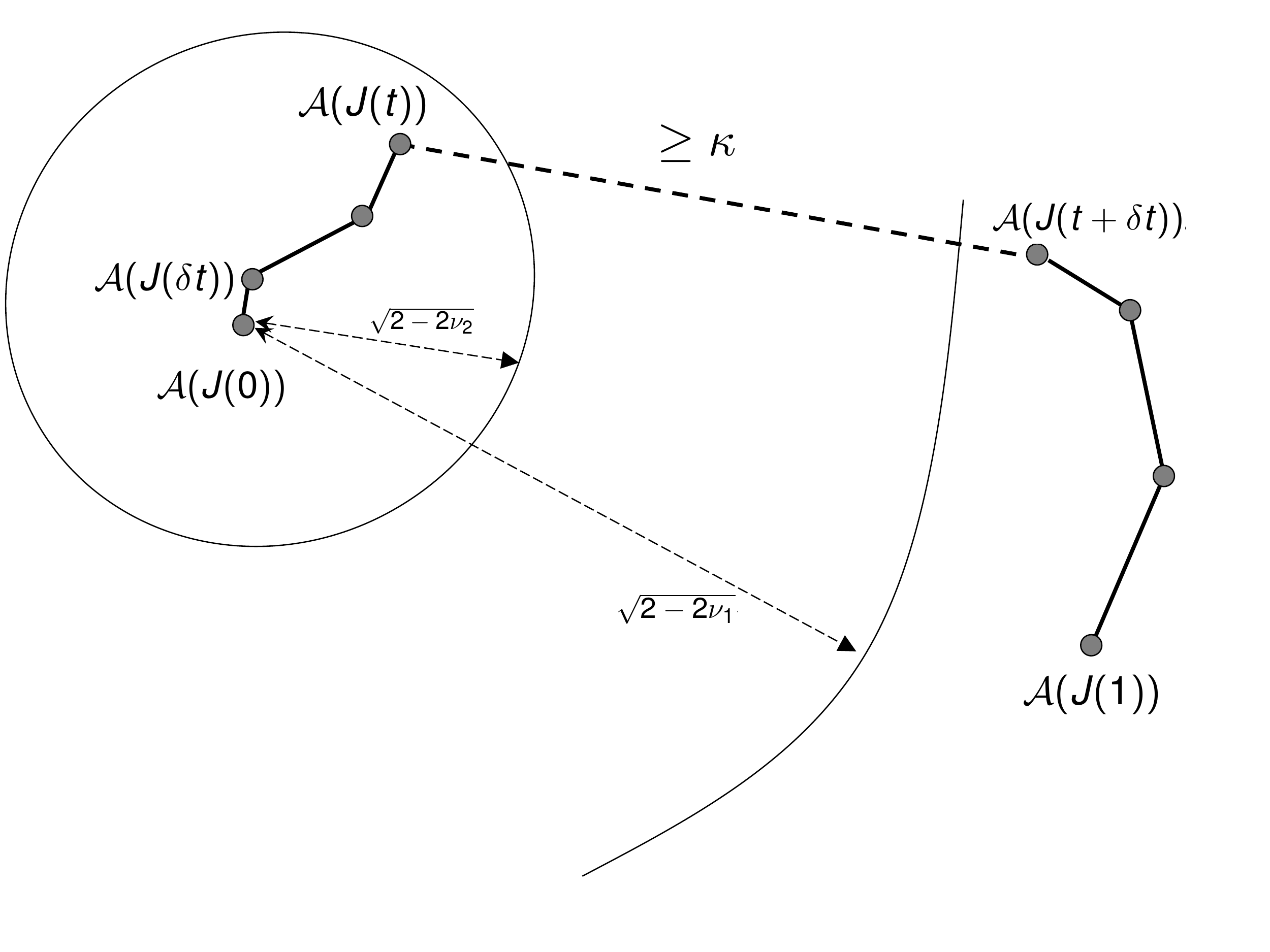}
\end{center}
\caption{The smaller circle represents $\eta_{\rm p,OGP}$-optimal solutions at distance $\le \sqrt{2-2\nu_2}$
from $\mathcal{A}(J(0))$. The complement to the larger circle 
represents $\eta_{\rm p,OGP}$-optimal solutions at distance $\ge \sqrt{2-2\nu_2}$
from $\mathcal{A}(J(0))$. As distance between the circle boundaries is $\sqrt{2-2\nu_1}-\sqrt{2-2\nu_2}\triangleq \kappa$,
at some instance $t$ the distance between ``successive'' solutions 
$\mathcal{A}(J(t))$ and $\mathcal{A}(J(t+\delta t))$ has to be at least $\kappa$, contradicting stability.
}
\label{fig:Stable-algorithms}
\end{figure}

The proof above is just an outline of the main ideas that have  different specific implementations for specific problems. The earliest application of this idea
was in~\cite{gamarnik2014limits}, in a different context of finding large independent sets in sparse random graphs. 
The method was used to show that local algorithms, appropriately defined, are stable, where $J$ denotes random graph connectivities. 
In the context of spin glasses, it was shown 
in~\cite{gamarnik2021overlapAukosh} that the AMP algorithm is stable and thus cannot overcome $\eta_{\rm p,OGP}$ barrier.
This was generalized in~\cite{gamarnik2020lowFOCS} where algorithms based on low-degree polynomials were shown to be stable. In the same paper Langevin dynamics was shown to be stable for spherical spin models when the running time is linear in $N$.
Extending the limitation of the Langevin dynamics beyond linear bound is an interesting open problem. A natural conjecture
is that the Langevin dynamics  produces a value at most $\eta_{\rm p,OGP}$ when run for $N^{O(1)}$ time.

By leveraging the multi-e-OGP method, which involves studying overlap patterns of more than two solutions, 
the barrier $\eta_{\rm p,OGP}$ and its analogues for other models
can be pushed to the value achievable by the state of the art algorithms. These algorithms are  
AMP in the $p$-spin Ising case~\cite{el2021optimization} and the spherical $p$-spin model case~\cite{subag2021following},
simple greedy algorithms for the case of random $K$-SAT problem and
the case of independent sets in sparse random graphs. 
The implementation of the multi-e-OGP for spin glass models was done
by Huang and Sellke~\cite{huang2021tight}, who have implemented a very ingenious version of the multi-OGP,  called branching-OGP. This version 
was motivated by the ultrametric structure of the solution space of $p$-spin models, widely conjectured to hold.
The implementation for the random K-SAT was done by Bresler and Huang~\cite{bresler2021algorithmicFOCS2021},
and  for independent sets in sparse random graphs by Wein in~\cite{wein2020optimal}.

Arguably the strongest implication of the OGP as an algorithmic barrier is its usage for establishing the state of the art lower bounds on depth of polynomial
size Boolean circuits. There is a long history in the theoretical computer science literature on establishing such lower bounds for various problems. 
In the context of constraint satisfaction problems, the prior state of the art result was achieved by 
Rossman~\cite{rossman2008constant,rossman2010average}
(see also extensions in~\cite{li2017ac,rossman2018lower}),  
who established
a depth lower bound $\Theta(\log n/(\kappa_n\log\log n))$ for poly-size circuits deciding the presence of an independent set of size $k_n$ in graphs with $n$ nodes. When the depth of the circuit is bounded by an $n$-independent constant, he showed that the size of the circuit has to be at least $n^{\Omega(\log n)}$.
This was done in the regime of random graphs where the typical value of $k_n$ grows at most logarithmically in $n$. Using the OGP method this bound was improved
to $\Theta(\log n/\log\log n)$, though for the search as opposed to the decision problem~\cite{gamarnik2020hardness}. Similarly, 
when the depth of the circuit is at most a constant, a stretched exponential lower bound $\exp(n^{\Omega(1)})$ on the size was established as well. 
It is in the context of this problem where the concentration around expectation adopted in the proof sketch does not hold, and furthermore, the stability
property does not hold w.h.p. Instead the idea  was to establish that circuits with small depth have stability property
with \emph{at least} sub-exponentially small probability. On the other hand, the stability can occur only for 
the event which is complementary to the OGP, and this complement event holds with exponentially small probability, thus leading to a contradiction.

A similar application of the OGP based method
shows that poly-size circuits producing solutions larger than $\eta_{\rm p,OGP}$ in $p$-spin models also have depth at least $\Theta(\log n/\log\log n)$. 
Pushing this result towards the value algorithmically achievable by the AMP, say using the Huang and Sellke~\cite{huang2021tight} is not immediate
due to the overlap Lipschitz concentration assumption required in~\cite{huang2021tight}. This extension is an interesting open problem.

Broadly speaking a big outstanding challenge is the 
applicability of OGP or similar methods for models with a planted signal, which we discuss in the following sections. While a version of 
OGP takes place in many such models, its algorithmic implication is far narrower than in the settings discussed above, such as $p$-spin models
and random constraint satisfaction problems. This presents an interesting and rather non-trivial challenge for future. 

\subsection{Connections with Replica Symmetry, Symmetry Breaking and the clustering (shattering) property}
\label{subsec:RSB-and-Clustering}
We discuss these connections rather informally now, leaving the technical aspects to other sources which we reference here.

The OGP arose in connection with studying the replica symmetry, replica symmetry breaking and related properties of spin glasses and their variants. 
Specifically, it arose as a method of   proving that  the set of satisfying 
solutions of a random constraint satisfaction problem is clustered (sometimes called shattered), meaning that it can be partitioned into ``connected'' components with order $\Theta(N)$ distance between them. How can one establish the existence of such a clustering picture? If the model exhibits the OGP say with parameters $\nu_1<\nu_2$, then clustering follows immediately, provided that solutions at distances $\sqrt{2-2\nu_1}$ or larger exist, as in this case one defines clusters as the set of solutions which can be reached from each other by paths in the underlying Hamming cube. The fact that distances between  $\sqrt{2-2\nu_2}$ and $\sqrt{2-2\nu_1}$ do not exist between the pairs of solutions imply that at least two (but in fact many) clusters exist. 

There are several caveats associated with this connection between the OGP and the clustering property. First this connection is one directional, in the sense that the presence of clustering does not necessarily imply the OGP, for a very simple reason: the diameter of the cluster can in principle be larger than the distances between the clusters. In this case, while the clustering property takes place, the set of all normalized pairwise distances could potentially span the entire interval $[0,1]$ without any gaps. Therefore the path towards establishing algorithmic lower bounds is not entirely clear.

Second, as it turns out in some models and in some regimes, the clustering picture has been established for the ``majority'' of the solution space, and not for the entire solution space. We will call it the weak clustering property, to contrast with the strong clustering property, which refers to a clustering property without exceptions.
For example, for the random K-SAT problem the onset of the  clustering property is known to take place  close to the threshold $(2^K/K)\log K$ for the clauses to variables densities, when $K$ is large, but only in the weak clustering sense discussed above: most but not necessarily all of the solutions can be split into clusters~\cite{AchlioptasCojaOghlanRicciTersenghi}. 

As it turns out, these exceptions are not just a minor nuisance, and can have profound algorithmic implications. The so-called symmetric perceptron model is a good demonstration of this~\cite{aubin2019storage,abbe2021proof,abbe2021binary,gamarnikBinaryPerceptron,perkins2021frozen}. For this model, the weak clustering property is known to take place at all constraints to variables densities, yet polynomial time algorithms exist at some strictly positive density values~\cite{abbe2021binary}. The multi-OGP analysis conducted in~\cite{gamarnikBinaryPerceptron} reveals that the gaps in the overlaps occur at densities \emph{higher} than the known algorithmic thresholds and thus the thresholds for the weak clustering property and the OGP do not coincide and, furthermore, the weak clustering property is apparently not a signature of an algorithmic hardness. Whether the strong clustering property can be used as a ``direct'' evidence of algorithmic hardness remains to be seen. For the further  discussion of the connection between the OGP, the weak and strong clustering properties, and the algorithmic ramifications, we refer the reader to~\cite{gamarnik2021overlap}.

Next we discuss the connection between the OGP, replica symmetry, symmetry breaking and the Parisi measure $\mu$. The Parisi measure $\mu$ arises in studying the Gibbs measure associated
with Hamiltonian $H$. (Very) roughly speaking, it describes an overlap structure of two
nearly optimal solutions $\sigma$ and $\tau$ chosen uniformly at random. This can be formalized
by introducing a small positive temperature parameter in the Gibbs distribution, but
we skip this formalism. The idea is that $(1/N)|\langle \sigma,\tau\rangle|$ has the Cumulative Distribution
Function (CDF) described by $\mu$ in the large $N$ limit. The support of $\mu$ is naturally some subset of $[0,1]$.
The source of randomness is dual here, one arising from the randomness
of the Hamiltonians, and one arising from the sampling procedure. Whether $\mu$ is indeed the 
limit the CDF of the overlaps in the limit remains a conjecture, which has been confirmed only for the spherical case. Loosely speaking
the model is defined to be in the replica symmetric regime (RS) if $\mu$ is just a $\delta$ mass
at zero. Namely, the overlap $(1/N)\langle \sigma,\tau\rangle$ is approximately zero with high 
probability, implying that  typical pairs of solutions are nearly orthogonal to each other. 

Replica symmetry breaking (RSB) then refers to $\mu$ being distinct from this 
singleton structure. Now if the model exhibits OGP, then a part of
$\mu$ is flat: the CDF of the overlaps is constant on $(\nu_1,\nu_2)$. Namely, the CDF \emph{is not} strictly increasing. 
The absence of this flat part of $\mu$ is exactly what was used  in constructions of near optimal solutions in~~\cite{subag2021following,montanari2021optimization,el2021optimization},
(and the presence of the OGP is an algorithmic obstruction as we have discussed). 
So presumably, we could have used the flatness of the Parisi measure as
a ``certificate'' of hardness. However, there are challenges associated with this
alternative. First, as we have discussed, whether $\mu$ indeed describes the distribution
of overlaps remains an open question, whereas the presence of the OGP has been confirmed. More importantly though, even
modulo the $\mu$ being the accurate descriptor of the overlaps,
the connection between OGP and the flatness
of $\mu$ is one-directional. The flatness of $\mu$ in some intervals $(\nu_1,\nu_2)$
means only that the density of the overlaps falling into this interval is asymptotically
zero after taking $N$ to infinity. It does not imply the absence of such overlaps. 
This is similar to the distinction between the weak and strong clustering property:
most of the overlaps are outside of the flat parts, but exceptions might exist.
The presence of such exceptions is bad news for the efforts of establishing 
algorithmic lower bounds. Not only the argument for proving the algorithmic lower
bounds appears to break down, but
also the presence of exceptions, namely a small number of overlaps falling into
this interval, might be potentially a game changer, as we saw in the case
of the symmetric perceptron model. 





\section{Statistical and computational trade-offs in inference and learning}\label{sec:tradeoffs}

In this section we move from optimization problems to statistical inference, in other words from the non-planted problems to the planted ones. We recall our working examples defined in section \ref{sec:problem_defs}, that cover a large range of settings and applications, the spiked tensor model and the generalized linear model. 

In order to describe the conjectured results on the algorithmic hardness of the planted problems we will first discuss the Bayes-optimal inference of the planted configuration from observations. We will then show how to analyze the performance of the Bayes-optimal inference in the large size limit $N\to \infty$ and under the stated randomness of the generative model. We will then show that phase transitions in the capability of the Bayes-optimal estimator to reconstruct the signal have an intriguing algorithmic role as a suitable type of message passing algorithms are able to reach optimal performance for all parameters except in the metastable region of first order phase transitions. This metastable region is then conjectured to be algorithmically hard -- the hard phase. Section \ref{sec:sos} will then present the currently strongest known method for showing evidence of such hardness in some cases. 

\subsection{The minimum mean-squared error} 

In both the spiked tensor model and the generalized linear model as defined in section \ref{sec:problem_defs} the optimal inference of the planted signal $u$ can be achieved by computing the marginals of the posterior probability distribution 
\begin{align}
    P(z| Y) = \frac{1}{{\cal Z}} P(z) P(Y|z) \, . 
\end{align}
Concretely, when aiming to find an estimator $\hat z$ that would minimize the mean-squared error to the signal $u$
\begin{align}
    {\rm MSE}(\hat z) = \frac{1}{N} \sum_{i=1}^N (u_i - \hat z_i)^2 
\end{align}
 we conclude that from all the possible estimators we should take $\hat z$ to be the marginal of the posterior
 \begin{align}
        \hat z_i = \mathbb{E}_{P(z|Y)}(z_i) \, .
 \end{align}
We will call the MSE achieved by this estimator the minimum-MSE, abbreviated MMSE. 
In the large size limit $N\to \infty$ computing marginals over $P(z|Y)$ with $z\in \mathbb{R}^N$ is in general exponentially costly in $N$, and thus potentially computationally hard even in the specific probabilistic generative models from Section~\ref{sec:problem_defs}. 

However, for the spiked tensor model as well as for the generalized linear model tools from the theory of spin glasses come to the rescue and allow us to analyze the value 
of the MMSE in the larger size limit as well as design message passing algorithms with properties closely related to the approach to obtain the MMSE. Let us start by describing the form in which we obtain the asymptotic value of the MMSE. Replica theory allows us to derive an explicit formula for a function $\Phi_{RS}(m)$, $m\in \mathbb R$, called the replica symmetric free entropy such that 
\begin{align}
    \lim_{N\to \infty} \mathbb{E}_{Y,u,J} \log{\cal{Z}} = \max_m \Phi_{RS}(m)\, .
    \label{eq:phi_RS}
\end{align}
We note that in physics it is more common to define the free energy which is just the negative of the free entropy. 
The average over $Y,u,J$ applies to the generalized linear model. In the spiked matrix model the $Y$ can be dropped as in the definition we gave it explicitly depends on $u$ and $J$. The function $\Phi_{RS}(m)$ explicitly depends on the parameters of the prior, the likelihood and the ratio $\alpha = N/P$, but in our notation we omit this dependence. 
We then call 
\begin{align}
    m^* = {\rm argmax} \, \Phi_{RS}(m)
\end{align}
and state a generic result for the MMSE that is given by the global maximizer of the replica symmetric free entropy
\begin{align}
    \lim_{N\to \infty} {\rm MMSE} = \rho - m^* 
\end{align}
where the constant $\rho = \mathbb{E}(u_i^2)$ is simply the second moment of the signal components. 

The derivations of these result and the explicit formulas for $\Phi_{RS}(m)$ were given in the spin glass literature for many special cases and mostly without a rigorous justification. In the general form considered in this paper and including rigorous proofs they were given for the spiked tensor model in \cite{lesieur2017statistical}, and  for the generalized linear model in \cite{barbier2019optimal}. For the purpose of this paper we will stay on the abstract level expressed above because on this level the discussion applies to a broad range of settings and we do not want to obfuscate it with with setting-dependent details. 

An important comment needs to be made here about the very generic validity of the replica symmetric result for the free entropy in the Bayes-optimal setting, i.e. when the prior and likelihood match the corresponding distributions in the model that generated the data. By the very nature of the Bayes' formula the signal $u$ has properties interchangeable with properties of a random sample from the posterior $P(z|Y)$. This is true even at finite size $N$ and even for models where $J$ is not random and where the likelihood and the prior are not separable. A consequence of the interchangeability is that under the averages over the posterior measure and the signal $u$ we can replace the signal $u$ for a random sample from the posterior and vice versa. This is called the Nishimori condition in the statistical physics literature \cite{nishimori2001statistical,zdeborova2016statistical}. A direct consequence of the Nishimori condition is that the magnetization (correlation between the signal and a random sample) and the overlap (correlation of two random samples) have to be equal, which in return means that the overlap distribution needs to be concentrated on a delta function and thus no replica symmetry breaking is possible in the Bayes-optimal setting. The Nishimori conditions also play a key role in the proof techniques used to establish the above results rigorously in \cite{lesieur2017statistical,barbier2019optimal}.

It it also important to note that what we discuss in this section is limited to the large size limit $N\to \infty$ with parameters scaling in such a way with $N$ for the MMSE to go from $\rho$ to $0$ as the signal-to-noise ratio $\alpha$ increases from 0 to large $O(1)$ values. This imposes scaling on the $\lambda_N$ for the spiked tensor model that is $O(N^{(1-p)/2})$. This will be in particular important for our claims about the optimality of the AMP algorithm that will be restricted to this regime and will not necessarily apply to performance of AMP for much larger signal to noise ratios. 

\subsection{AMP and its state evolution}

In the previous section we analyzed the MMSE as it would be achieved by the exact computation of the posterior average. This is, however, in general computationally demanding and thus a next natural question is whether we can reach this MMSE computationally efficiently. 
Message passing algorithms provide an algorithmic counter-part of the replica method. In particular, the approximate message passing algorithm (AMP) that is an extension of the TAP equations \cite{thouless1977solution} to the general setting of the spiked tensor model and the generalized linear model is of interest to us in this paper. AMP is an iterative algorithm that aims to compute the Bayes-optimal estimator~$\hat z$. Schematically the update of AMP at time step $t$ for the AMP's estimate $z^{t}_{\rm AMP} \in \mathbb{R}^N$ can be written for both the considered models as 
\begin{align}
     z^{t+1}_{\rm AMP} = {\cal F}(z^{t}_{\rm AMP}) \,  
\end{align}
for an update function ${\cal F}(.)$ that depends on $Y$, parameters of the prior and the likelihood, and for the generalized linear model also on $J$. 

The key property that makes AMP so theoretically attractive is that in the large size limit the accuracy of the AMP estimator can be tracked via low-dimensional set of equations called state evolution. To state this we introduce the correlation between AMP estimate and the signal at iteration~$t$
\begin{align}
    m_N^t = \frac{1}{N} \sum_{i=1}^N u_i\,  (z^{t}_{\rm AMP})_i
 \end{align}
The state evolution implies that this quantity in the large size limit $m^t = \lim_{N\to \infty} m_N^t$ behaves as
\begin{align}
        m^{t+1} = f_{\rm SE} (m^{t})\, , 
        \label{eq:SE}
\end{align}
for a function $f_{\rm SE}$ that depends on the parameters of the models, but not any longer of any high-dimensional quantity. 
The state evolution of AMP is a crucial contribution that came from mathematical developments of the theory \cite{bolthausen2014iterative,bayati2011dynamics} and was not known in its current form in the statistical physics literature before that. The proofs of state evolution have been extended to a broader setting \cite{javanmard2013state,bayati2015universality,gerbelot2021graph}.

What makes the state evolution particularly appealing in the statistical physics context is its connection to the computation of the MMSE. The fixed points of the expression (\ref{eq:SE}) can be expressed at the stationary points of the replica symmetric free entropy
\begin{align}
      m =  f_{\rm SE} (m) \quad \Leftrightarrow \quad \frac{\partial \Phi_{\rm RS}(m)}{\partial m} = 0
\end{align}
where $\Phi_{\rm RS}(m)$ is indeed the same free entropy as in eq.~(\ref{eq:phi_RS}). 

Since the signal $u$ is unknown the corresponding initialization is $m^{t=0}=0$ (this is for prior distribution with zero mean) and thus the performance of AMP is given by the stationary point of the free entropy that is reached by iterating (\ref{eq:SE}) initialized at $m^{t=0}=0$. The performance of AMP at convergence thus corresponds to the local maximum $m_{\rm AMP}$ of the free entropy $\Phi_{\rm RS}(m)$ that has the largest error. The corresponding MSE is then 
\begin{align}
    {\rm MSE}_{\rm AMP} = \rho - m_{\rm AMP}\, . 
\end{align}

\subsection{The phase diagrams and the hard phase}

We have seen in the previous two subsections that the values of the MMSE as well as the MSE obtained by the AMP algorithm can both be deduced from the extremizers of the free entropy function $\Phi_{\rm RS}(m)$. 

While the MMSE is given by the global maximizer of $\Phi_{RS}(m)$, the MSE reached by the AMP algorithm is given by the maximizer having the smallest $m$. In the following we will consider all the extremizers of $\Phi_{RS}(m)$ as this will allow us to understand the resulting overall picture. We will discuss how the extremizers depend on some kind of signal to noise ratio $\alpha$. This signal to noise ratio can be simply the value of $\alpha=\lambda$ in the spiked matrix model, or the sample complexity ratio $\alpha = P/N$ in the generalized linear model. 

Depending on the other parameters of the model we can observe a number of scenarios, we will discuss several of them below and refer to examples where they appear. In the following sketches all the colored curves are extremizers of $\Phi_{RS}(m)$. Those in {\color{blue}{blue}} are the global maximizers of the free entropy corresponding to the MMSE. No algorithmic procedure can achieve an error lower than the MMSE. When the AMP algorithm does not achieve the MMSE, the MSE it reaches at its fixed point corresponds to a maximizer of the free entropy of a higher error ${\rm MSE}_{\rm AMP}$ depicted in {\color{green}{green}}. In {\color{red}{red}} we depict the other extremizers of the free entropy, in dashed red the minimizers, and in full red the other maximizers. 

The region of error between the green and the blue curve are values of the MSE that are information-theoretically reachable, but the AMP algorithm does not reach them. We call this region the \textbf{hard phase}, and its boundaries on the signal-to-noise ratio axes: $\alpha_{\rm IT}$ for the information theoretic threshold where the values of the two maximizers of $\Phi_{\rm RS}(m)$ switch order, and $\alpha_{\rm alg}$ above which AMP reaches the MMSE. The hard phase exists in between these two thresholds, $\alpha_{\rm IT} < \alpha < \alpha_{\rm alg}$. A third threshold $\alpha_s$ marks the spinodal point at which the lower-error maximizer of the free entropy ceases to exist, this point does not have significant algorithmic consequences for finding the signal. In other cases there may be no phase transition at all or a second order (continuous) phase transition marked by $\alpha_c$. 

The physical interpretation of the cases where the hard phase exists is the one of first order phase transition in a high-dimensional (mean-field) system. The $\alpha_{\rm IT}$ corresponds to the thermodynamic phase transition while $\alpha_s$ and $\alpha_{\rm alg}$ are the spinodals, i.e. the boundaries of the metastable regions. In the hard phase the thermodynamic equilibrium corresponds to the higher free entropy branch depicted in blue, and the green fixed point corresponds to the metastable state. In the region $\alpha_{s} < \alpha < \alpha_{\rm IT}$ the AMP algorithm finds the thermodynamic equilibrium, but this state is split into exponentially many separated states, each corresponding to the metastable branch (full red). In the language of replica-symmetry breaking this phase corresponds to the dynamical-1RSB phase (d-1RSB). In the d-1RSB phase the AMP algorithm reached optimal performance in terms of finding the signal, however, sampling the posterior measure in the d-1RSB region is conjectured computationally hard.

\begin{figure}[!ht]
\vspace{-1.5cm}
 \includegraphics[width=150mm]{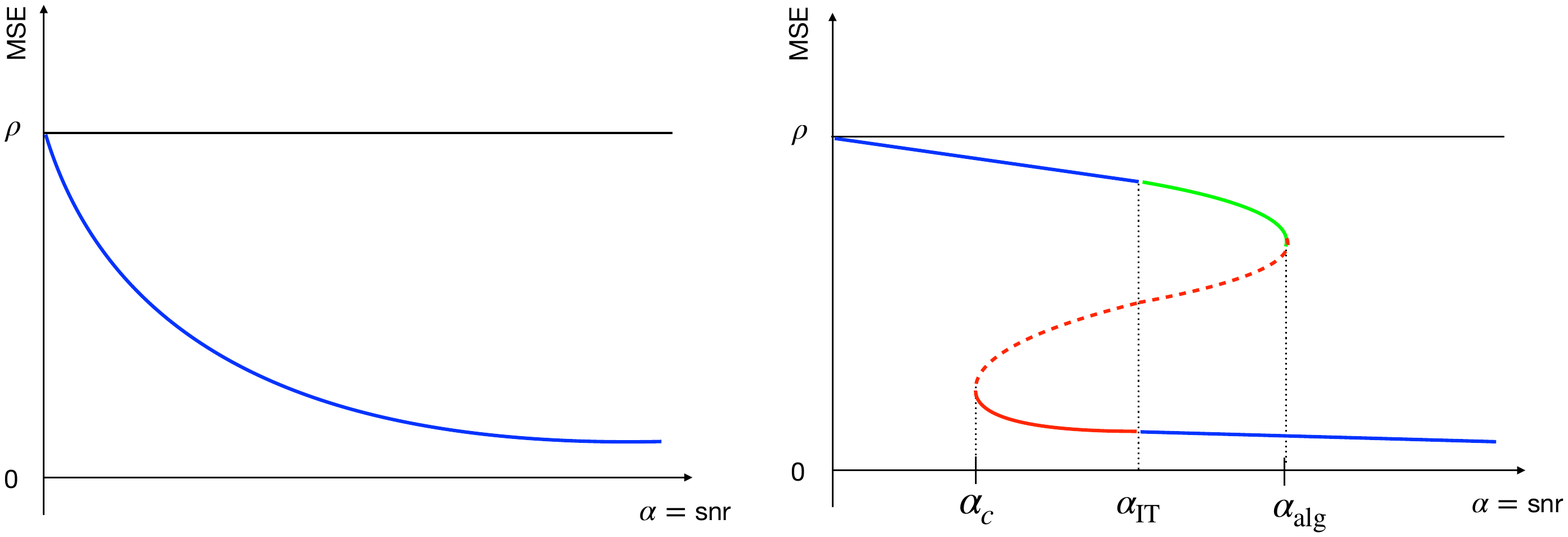}
\vspace{-2cm}
 \caption{Extremizers of the replica symmetric free entropy when neither $m=0$ nor $m=\rho$ are stationary points. Colors explained in the text. (Left) A case without a phase transition. (Right) A case with a first order phase transition.}
 \label{fig:phase_basic}
\end{figure}

In Fig.~\ref{fig:phase_basic} we depict one possible structure of extremizers of the free entropy $\Phi_{\rm RS}(m)$ for models where neither $m=0$ nor $m=\rho$ are fixed points for $\alpha>0$. On the left hand side of Fig.~\ref{fig:phase_basic} we depict a case without a phase transition. This situation arises for instance in generalizes linear models with Gaussian prior and a sign activation function, corresponding to the spherical teacher-student perceptron, see e.g. center of Fig.~2 in \cite{barbier2019optimal} for a concrete example. 
On the right hand side of Fig.~\ref{fig:phase_basic} we depict a case with a first order phase transitions. Such as situation arises for instance spiked matrix model where the prior is sparse with non-zero mean, see e.g. rhs of Fig.~4 in \cite{lesieur2017constrained} for a concrete example. 

\begin{figure}[!ht]
\vspace{-1.5cm}
 \includegraphics[width=150mm]{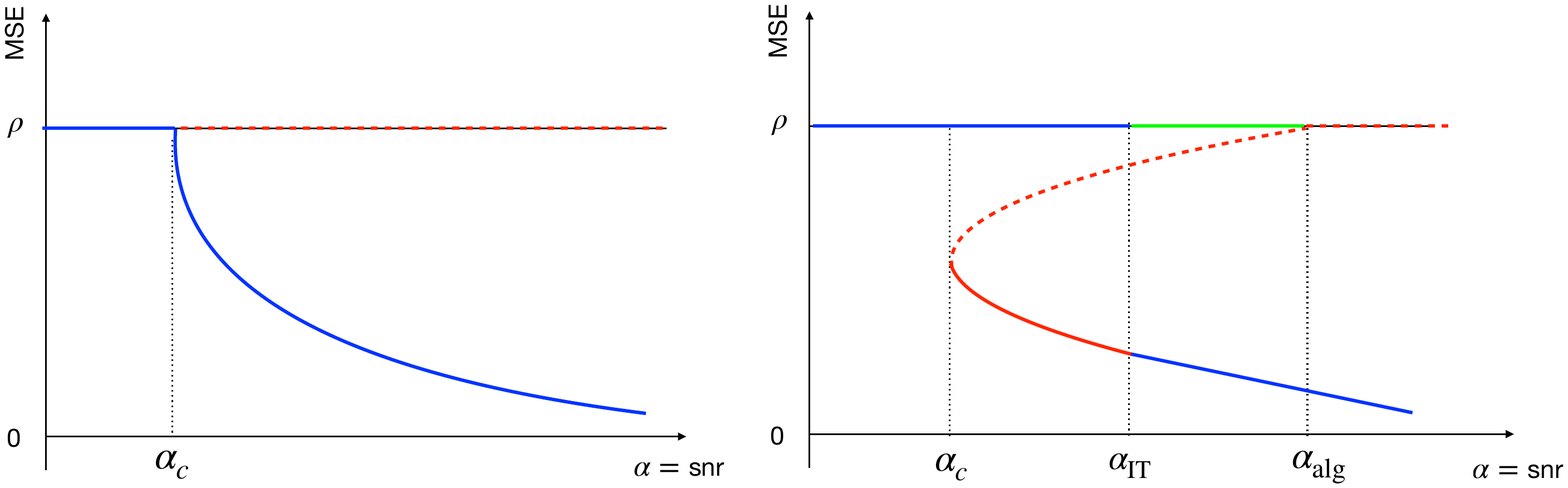}
\vspace{-2cm}
 \caption{Extremizers of the replica symmetric free entropy when $m=0$ is a stationary point for all $\alpha$. Colors explained in the text. (Left) A case with a (continuous) 2nd order phase transition. (Right) A case with a (discontinuous) first order phase transition.}
 \label{fig:phase_detect}
\end{figure}

In Fig.~\ref{fig:phase_detect} we depict another possible structure of extremizers of the free entropy $\Phi_{\rm RS}(m)$ for models where $m=0$ is a fixed point. On the left of Fig.~\ref{fig:phase_detect} there is a situation with a second order phase transition as is the case for instance in the symmetric stochastic block model with two groups, see e.g. Fig.~1 in \cite{decelle2011asymptotic} for a specific example. On the right of Fig.~\ref{fig:phase_detect} there is a situation with a first order phase transition as is the case for instance in the symmetric stochastic block model with more than 4 groups, see e.g. Fig.~3 in \cite{decelle2011asymptotic} for a specific example. In this case the threshold at which the fixed point at $m=0$ ceases to be a maximum and start to be a minimum is the well-known Kesten-Stigum threshold \cite{kesten1966limit}, marked $\alpha_c$ on the lhs of the figure, and $\alpha_{\rm alg}$ on the rhs of the figure. When $m=0$ and ${\rm MMSE}=\rho$ is the thermodynamic equilibrium no correlation with the signal can be obtained and the phase $\alpha<\alpha_{\rm IT}$ is in this case referred to as the undetectable region. 
In this phase the planted model is contiguous to the non-planted model in the sense that all high-probability properties in the planted model are the same in the non-planted one other \cite{mossel2012stochastic}. 
This is the setting that is most often explored in the sum-of-squares approach of section \ref{sec:sos}.

\begin{figure}[!ht]
\vspace{-1.5cm}
 \includegraphics[width=150mm]{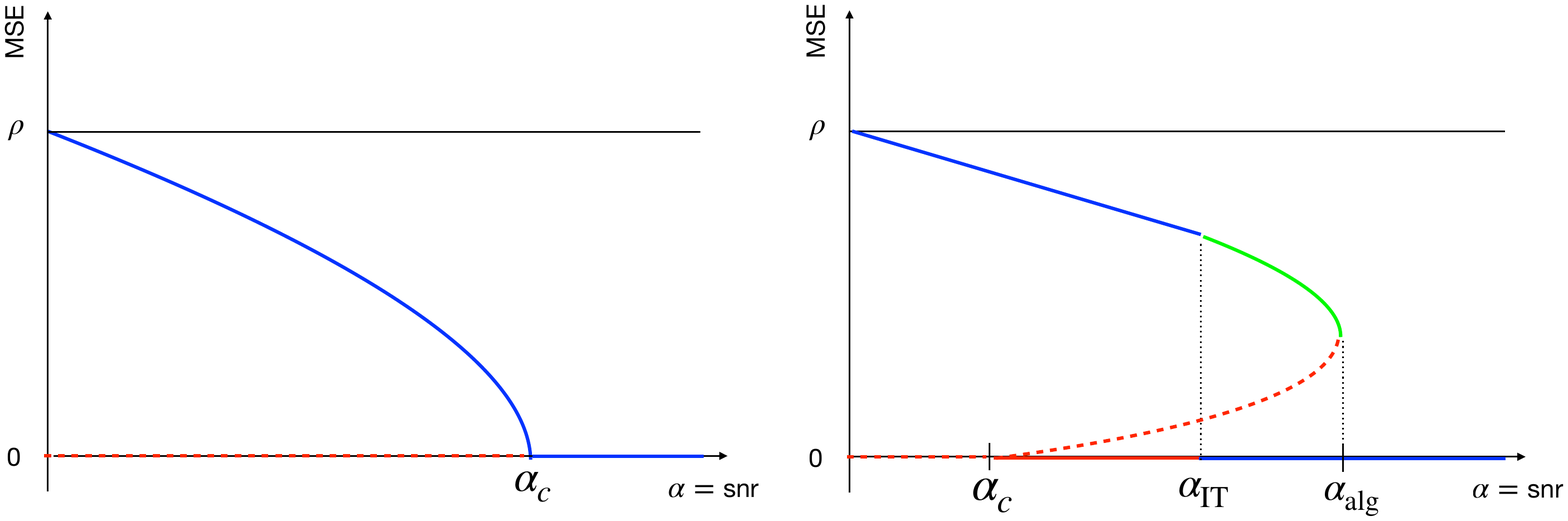}
\vspace{-2cm}
 \caption{Extremizers of the replica symmetric free entropy when $m=\rho$ is a stationary point for all $\alpha$. Colors explained in the text. (Left) A case with a 2nd order phase transition. (Right) A case with a first order phase transition.}
 \label{fig:phase_exact}
\end{figure}

In Fig.~\ref{fig:phase_exact} we depict yet another possible structure of extremizers of the free entropy $\Phi_{\rm RS}(m)$ for models where $m=\rho$ is a fixed point and thus where exact recovery of the signal with ${\rm MMSE}=0$ is  possible for sufficiently large signal-to-noise ratios. On the right of Fig.~\ref{fig:phase_exact} we depict a case with a first order phase transition. Such a situation arises e.g. in the generalized linear model with binary prior and sign activations, corresponding to the teacher-student binary perceptron, see left hand side of Fig.~2 in \cite{barbier2019optimal}. On the left of Fig.~\ref{fig:phase_exact} we depict a case with a second order phase transition, this arises e.g. in the generalized linear model with Laplace prior and no noise, corresponding to the minimization of the $\ell_1$ regularization, see e.g. Fig.~3 in \cite{krzakala2012probabilistic}.

The examples we depict in this section do not exhaust all the possible scenarios one encounters in computational problems. Some of those we did not cover include the planted locked constraint satisfaction problems where both $m=0$ and $m=\rho$ fixed points exist and an all-to-nothing first order phase transition happens between these two fixed points \cite{zdeborova2011quiet}. Both $m=0$ and $m=\rho$ fixed point also exist for instance in the generalized linear model with Gaussian prior and absolute value activation corresponding to the phase retrieval problem. In that case there is a second order phase transition from the undetectable phase to a detectable one and later on a first order phase transition to exact recovery, see e.g. left hand side of Fig.~5 in \cite{barbier2019optimal}.

Another interesting and very generic case is depicted e.g. in Fig.~6 of \cite{lesieur2017constrained} for the spiked matrix model with a symmetric Rademacher-Bernoulli prior. In this case the undetectable phase ($m=0$ fixed point) is followed by a phase where a correlation with the signal is detectable but small, and where AMP reaches a small but suboptimal correlation to the signal. The position of the first order phase transition can be either before or after the detectability threshold (as in the left or right of the lower part of Fig.~6 in \cite{lesieur2017constrained}). While this may seem a rare scenario, results in \cite{ricci2019typology} (see Fig.~2) actually indicate that it is likely very generic and that often the size of the region where detection is possible but sub-optimal is very thin.  

Yet another interesting example of a phase transition in a planted problem is the planted matching problem where the phase transition is infinite order, i.e. all the derivatives of the order parameter $m$ exist at the transition from partial recovery phase to exact recovery phase \cite{semerjian2020recovery}. 

\subsection{Is the hard phase really hard?}
\label{sec:is_it_hard}

A fundamental question motivating the discussion of this paper is for what class of algorithms is the hard phase computationally inaccessible? 

An important evidence towards the hardness is summarized in 
\cite{celentano2020estimation} where it is shown that a very broad range of algorithms related structurally to the approximate message passing cannot improve over the AMP that uses the Bayes-optimal parameters. Efforts to prove lower bounds are considerable, as discussed in section \ref{sec:sos}. A number of authors put forward a conjecture that in settings where the large-size limit and randomness is taken in such a way that AMP and the Bayes-optimal solution are related in the way we describe above, then AMP is optimal among a large class of algorithms. But could this possibly be all polynomial algorithms? 

It is important to note that there are problems with the phenomenology leading to the hard phase yet for which polynomial algorithms to find the signal exist never-the-less. One of them is that planted XOR-SAT problem \cite{franz2001ferromagnet,zdeborova2011quiet} that is mathematically a linear problem in the Boolean algebra and can thus always be solved using Gaussian elimination. Gaussian elimination, however, runs with time larger than linear in the size of the system and is not robust  to noise where we plant a solution that violates a small fraction of clauses. A more surprising and recent example is given by the noise-less phase retrieval problem for Gaussian matrix~$J$ where the so-called LLL algorithm also works in polynomial time down to the information-theoretic threshold \cite{gamarnik2021inference,song2021cryptographic}. The phase retrieval problem is NP-hard, unlike the planted XOR-SAT. Again the LLL is based on linear algebra and thus in some sense related to Gaussian eliminations, it is not robust to noise, or runs in time that is polynomial with an exponent considerably larger than one. 

The existence of these examples makes it clear that in some cases other algorithms can perform better than AMP with the Bayes-optimal parameters in the high-dimensional limit. It is thus more reasonable to conjecture that the AMP algorithm may be optimal among those polynomial ones that are required to be robust to noise?  Or among those that run with resources linear with the input size of the problem (i.e. quadratic in $N$)?

We also want to note here another case that is often cited as an example where other algorithms beat AMP. This is the  spiked tensor model  for $p\ge 3$. However, in this case the algorithmic threshold happens at $\lambda_N \sim N^{-p/4} $ while the information theoretic one at $\lambda_N \sim N^{(1-p)/2}$. We do not expect AMP to be in general optimal for other scalings than the information-theoretic one, we thus do not consider this as a counter-example to the conjecture of optimality of AMP. Our conjectures about optimality of AMP restrict to the information-theoretic scaling.



 

\subsection{The hard phase is glassy, causing hurdles to gradient-based algorithms}

From the physics point of view the conjecture of optimality of AMP is very intriguing. It needs to be stressed that the state evolution that rigorously tracks the performance of the AMP algorithm corresponds to the replica symmetric branch of the free entropy while replica symmetry breaking is needed to describe the physical properties of the metastable state \cite{antenucci2019glassy}. 

Physically, and following the success of survey propagation \cite{braunstein2005survey} in solving the random K-SAT problem, one may have hoped that including the glassiness in the form of the algorithm, as done in \cite{antenucci2019approximate}, would improve the performance. This is, however, not happening and is rigorously precluded by the proof of \cite{celentano2019fundamental}. So in a sense while AMP follows the non-physical solution for the metastable state, this solution has fundamental meaning in terms of being the best solution achievable by a computationally tractable algorithm. 

It is interesting to note that early work in statistical physics indeed dismissed the replica symmetric spinodal as non-physical, see \cite{sompolinsky1990learning}, and presumed that algorithms will be stopped by the glassiness of the metastable phase.  This is a nice example where the later state-evolution proof takes over the early physical intuition about what is the relevant algorithmic threshold. 

At the same time, the physics intuition of the glassiness stopping the dynamics for signal-to-noise ratios larger than where the replica symmetric appears was not wrong. It simply does not apply to the AMP algorithm that does not correspond to a physical dynamics as it does not perform a walk in the space of possible signals but rather iterates marginals over the signal components. If we consider now instead physical dynamics such and Monte-Carlo Markov chains (MCMC) or algorithms updating the signal estimate based on possibly noisy gradient descent the early intuition of \cite{sompolinsky1990learning} turned out to be completely correct in the sense that these algorithms actually perform considerably worse than AMP when the hard phase is present. Interestingly this was not expected in some works, e.g. \cite{decelle2011asymptotic} conjectured that MCMC performs as well as message passing in the stochastic block model, which turns out to be wrong \cite{chiara2022theory}. Very clear-cut examples of gradient-based Langevin algorithms performing worse than AMP are given for the mixed spiked matrix-tensor model in \cite{mannelli2020marvels} and for the phase retrieval in \cite{sarao2020complex}. 

The phase retrieval example is particularly relevant due to its interpretation as a neural network and given that gradient descent is the working horse of the current machine learning revolution. One may ask whether some key parts of the current machine learning tool-box such as over-parametrization and stochasticity in gradient descent are not a consequence of mitigation of the hurdles that gradient descent encounters due to glassiness of the landscape. Some resent works on the phase retrieval problem do point in that direction \cite{sarao2020optimization,mignacco2021stochasticity}.




\section{Polynomial Proofs: the Sum-of-Squares Hierarchy}
\label{sec:sos}

In the absence of a proof that $\mathrm{P} \ne \mathrm{NP}$, we have no hope of proving that problems in a certain parameter range truly require exponential time. There may in fact be no hard regimes. But we can try to gather the efficient algorithms we know of into large families---each characterized by a particular strategy or kind of reasoning, or which can only ``understand'' certain things about their input---and show that no algorithm in these families can succeed. In the previous section, we discussed how the overlap gap property can be used to defeat algorithms that are stable to noise or small perturbations in their input. 

Here we discuss classes of algorithms that have an algebraic flavor. We will focus on the sum-of-squares hierarchy, and briefly discuss its cousin the low-degree likelihood ratio. Many of the best algorithms we know of are captured by these classes, including powerful generalizations of spectral algorithms and classic approximation algorithms. Thus if we can show that they fail to solve certain problems, or more precisely that they require polynomial ``proofs'' or ``likelihood ratios'' of high degree, this constitutes additional evidence that these problems are hard. 

There are types of reasoning that these systems have difficulty with, as our first example will illustrate. This leaves open the possibility that some very different algorithm could efficiently solve problems in what we thought was a hard regime. However, these other types of reasoning seem fine-tuned and fragile, and only work in  noise-free settings. For a wide variety of noisy problems, algorithms associated with sum-of-squares are conjectured to be optimal~\cite{barak-steurer}.

\subsection{Proofs and refutations}

At its heart, the sum-of-squares (SoS) hierarchy is a way of constructing \emph{refutations} of constraint satisfaction or optimization problems: proofs that a solution does not exist, or that no solution  achieves a certain value of the objective function. It comes with a dual problem, of evading refutation by finding a \emph{pseudoexpectation}: a fictional distribution of solutions that looks reasonable as long as we only ask about polynomials up to a certain degree. If a pseudoexpectation can be constructed that ``fools'' polynomials up to degree $d$, then any refutation must have degree greater than $d$. 

Let's look at an example. Consider three variables $x,y,z \in \{\pm 1\}$. Is it possible for them to sum to zero? This problem may seem trivial, but bear with us. Algebraically, we are asking whether the following system of polynomials has a solution,
\begin{align}
\begin{array}{rl}
x^2-1 &\,=\, 0 \\
y^2-1 &\,=\, 0 \\
z^2-1 &\,=\, 0 \\
x+y+z &\,=\, 0 \, . 
\end{array} \label{eq:xyz-system}
\end{align}
Here is a proof, that the  motivated reader can verify, that no solution exists:
\begin{align}
\frac{1}{8} \Big[ & \big( x^2 + 3 (y^2 + z^2) + 4 (x y + x z + 3 y z) - 3 \big)
\,(x^2-1) \nonumber \\
+\; & \big( y^2 + 3 (x^2 + z^2) + 4 (yz + xy + 3 xz) - 3 \big)
\,(y^2-1) \nonumber \\
+\; & \big( z^2 + 3 (x^2 + y^2) + 4 (xz + yz + 3 xy) - 3 \big)
\,(z^2-1) \Big] \nonumber \\
+\; & (x+y+z)^2 \nonumber \\
& \qquad \qquad \qquad \qquad \qquad = \; \frac{1}{8} 
\left( (x+y+z)^2 - 1 \right)^2 + 1 \, .
\label{eq:xyz-proof}
\end{align}
If the constraints~\eqref{eq:xyz-system} hold, then the left-hand side of~\eqref{eq:xyz-proof} is identically zero. On the other hand, the right-hand side is the square of a polynomial plus $1$, giving the contradiction $0 \ge 1$. We will reveal below how we constructed this proof.

More generally, suppose we have a set of polynomials $f_1(\vec{x}),\ldots,f_k(\vec{x})$ over $n$ variables $x_1,\ldots,x_n$. We wish to prove that there is no $\vec{x} \in \R^n$ such that $f_i(\vec{x})=0$ for all $i$. A sum-of-squares proof consists of additional polynomials $g_1,\ldots,g_k$ and $h_1,\ldots,h_t$ such that 
\begin{equation}
\label{eq:sos-proof}
\sum_{i=1}^k g_i(\vec{x}) f_i(\vec{x}) 
= \sum_{j=1}^t h_j(\vec{x})^2 \;+\; 1 \, ,
\end{equation}
where $1$ on the right-hand side can be replaced by any positive constant. In other words, we find a linear combination of the $f_i$ that is strictly positive everywhere, so they can never be zero simultaneously. Any unsatisfiable system of polynomial equations $\{ f_i(\vec{x})=0 \}$ has a refutation of this form~\cite{krivine,stengle}. A logician would say that the SoS proof system is \emph{complete}. 


Now, we say a SoS proof is of degree $d$ if the polynomials $g_i f_i$ and $h_j^2$ on the left and right sides of~\eqref{eq:sos-proof} have maximum degree $d$. Thus our example~\eqref{eq:xyz-proof} is a proof of degree $d=4$. (By convention $d$ is always even: the $h_j$ have degree at most $d/2=2$.) As we increase $d$, we obtain a hierarchy of increasingly powerful proof systems. 

In some cases the  lowest possible degree of an SoS proof is much larger than the degree of the original constraints $f_i$, since we may need high-degree coefficients $g_i$ to create the right cancellations so that the sum can be written as a sum of squares. As we will see below, if the necessary degree grows with the size of the problem, we can interpret this as evidence that the problem is computationally hard.

\subsection{From proofs to algorithms: semidefinite programming}

Of course, the existence of an SoS proof doesn't necessarily make it easy to find. Algorithmically, how would we search for these polynomials? If we choose some ordering for the monomials up to some degree, writing a symbolic vector $\vec{m} = (1,x,y,z,x^2,xy,xz,y^2,\ldots)$, then we can represent a polynomial $q$ as a vector $\vec{q}$ of its coefficients and write $q(\vec{x})$ as an inner product $\inner{\vec{q}}{\vec{m}}$. Multiplying two polynomials is a bilinear operation, and the sum on the right-hand side of~\eqref{eq:sos-proof} can be written 
\begin{align}
\sum_{j=1}^t h_j(\vec{x})^2 
= \sum_{j=1}^t 
\inner{\vec{m}}{\vec{h}_j}
\inner{\vec{h}_j}{\vec{m}}
&= \bra{\vec{m}} \vec{H} 
\ket{\vec{m}} 
\nonumber \\
& \text{where} \quad 
\vec{H} = \sum_{j=1}^t \ket{\vec{h}_j} \bra{\vec{h}_j} \, . 
\label{eq:rhs-bilinear}
\end{align}
This bilinear form $\vec{H}$ is positive semidefinite, which we denote $\vec{H} \succeq 0$.


With this abstraction, the problem of finding SoS proofs asks for a positive semidefinite matrix that matches the left-hand side of~\eqref{eq:sos-proof}. To nail this down, for a polynomial $q$ let $q_u$ denote the coefficient of each monomial $u$. Then summing over all the cross-terms in the product of two polynomials $p, q$ gives
\begin{equation}
(pq)_u = \sum_{v,w:\,vw=u} p_v q_w \, . 
\end{equation}
Since for any two monomials $s,t$ the entry $H_{s,t} = \bra{s} \vec{H} \ket{t}$ must equal the coefficient of $u=st$ on the left-hand side of~\eqref{eq:sos-proof}, for any $s,t$ such that $st \ne 1$ we have
\begin{equation}
\sum_i \sum_{v,w:\,vw=st} (g_i)_s (f_i)_t 
= H_{s,t} 
\, ,  
\end{equation}
and for $s=t=1$ we have 
\begin{equation}
\sum_i (g_i)_1 (f_i)_1 = 1 + H_{1,1} \, . 
\end{equation}
For a given set $\{f_i\}$, these constraints are linear in the coefficients of the $\{g_i\}$. Adding the semidefiniteness constraint $\vec{H} \succeq 0$ to this linear system of equations makes this a case of \emph{semidefinite programming} or SDP~\cite{Shor-1987-SumOfSquares,Nesterov-2000-SOS,Parrilo-thesis,lasserre}.

SDP can be solved up to  arbitrarily small error in polynomial time whenever the number of constraints and the dimension of the matrices is polynomial. (There is an important caveat, namely that the coefficients of the SoS proof need to be polynomially bounded~\cite{ODonnell-2017-SOSNotAutomatizable,RW-2017-BitComplexity}.) Since the number of monomials over $n$ variables of degree $d$ is $\binom{n+d-1}{d} = O(n^d)$, this means that SoS proofs are easy to find whenever the degree $d$ is constant. 

On the other hand, if we can somehow prove that the lowest degree of any SoS proof grows with $n$, this rules out a large class of polynomial-time algorithms. When we can prove them, these SoS lower bounds are thus evidence of computational hardness.

\subsection{Sum-of-squares lower bounds: enter the Charlatan}

To see how we might prove such a lower bound, let's return to our earlier problem. A Charlatan\footnote{Many concepts in theoretical computer science have become personified over the years: the Adversary, the Oracle, Arthur and Merlin, Alice, Bob, and Eve, and so on. We propose that the Charlatan be added to this cast of characters.} comes along and claims that the system~\eqref{eq:xyz-system} has not just one solution, but many. That is, they claim to know a joint probability distribution over reals $x,y,z$ such that $x^2=y^2=z^2=1$ and $x+y+z=0$. To convince you, they offer to tell you the expectation $\Exp{q}$ of any polynomial $q(x,y,z)$ you desire---but only for $q$ of degree $d$ or less, where in this case $d=2$.

Let's call the Charlatan's claimed value for $\Exp{q}$ the \emph{pseudoexpectation}, and denote it $\Pse{q}$. How might you catch them in a lie? You are no fool; you know that the expectation of a sum is the sum of the expectations. Since the constraints $f_i(\vec{x})=0$ must hold identically, you also know that any $q$ that has $f_i$ as a factor must have zero expectation. Finally, you are well aware that the square of any polynomial is everywhere nonnegative, and thus has nonnegative expectation. 

Putting this together, the pseudoexpectation must be a linear operator from the space of polynomials of degree $d$ to $\R$ with the following properties:
\begin{enumerate}
    \item $\Pse{1}=1$
    \item $\Pse{f_i q}=0$ for any polynomial $q(x)$ of degree $d-\deg(f_i)$ or less
    \item $\Pse{q^2} \ge 0$ for any polynomial $q(x)$ of degree $d/2$ or less.
\end{enumerate}
Let's think of $\Psei$ as a bilinear form that takes two polynomials $p$, $q$ of degree up to $d/2$ and returns $\Pse{pq} = \bra{p} \Psei \ket{q}$. Then condition (3) corresponds to $\Psei$ being positive semidefinite, just as for $\vec{H}$ above. Since conditions (1) and (2) are linear, finding a pseudoexpectation is another case of semidefinite programming.

In our example, since $d=2$, the monomials that $\Psei$ needs to deal with are just $1, x, y, z$. Without further ado, we present the Charlatan's claim as a multiplication table of pseudoexpectations:
\begin{equation}
\label{eq:without-ado}
\begin{array}{c|cccc}
\Psei\, & \,1 & x & y & z \\ \hline
\,1\, & \,1\, & 0 & 0 & 0 \\
\,x\, & \,0\, & 1 & -1/2 & -1/2 \\
\,y\, & \,0\, & -1/2 & 1 & -1/2 \\
\,z\, & \,0\, & -1/2 & -1/2 & 1 
\end{array}
\end{equation}
That is, they claim that $x,y,z$ each have expectation $\Pse{x} = \bra{1} \Psei \ket{x} = 0$; they each have variance $\Pse{x^2} = \bra{x} \Psei \ket{x} = 1$; and each distinct pair is negatively correlated, with $\Pse{xy} = \bra{x} \Psei \ket{y} = -1/2$. As a result, $\Pse{x+y+z}=0$, and $\Pse{(x+y+z)p}=0$ for any linear function $p$, satisfying condition (2) above.

It is easy to check that this matrix of pseudomoments is positive semidefinite. Indeed its $3 \times 3$ part is the Gram matrix of three unit vectors that are $120^\circ$ apart. This is impossible for three real-valued variables in $\{\pm 1\}$, but as far as quadratic polynomials of $x,y,z$ are concerned, there is no contradiction. 

On the other hand, we already know that we can debunk the Charlatan's claims if we ask about degree-$4$ polynomials.  The left-hand side of~\eqref{eq:xyz-proof} must have zero expectation since it is a linear combination of the $f_i$. By linearity, this would imply that
\begin{equation}
\Pse{\frac{1}{8} 
\left( (x+y+z)^2 - 1 \right)^2} = -1 < 0 \, .
\end{equation} 
Thus there is no way to extend the pseudoexpectation in~\eqref{eq:without-ado} from degree 2 to degree 4 without violating positive semidefiniteness. 
More generally, an SoS proof of the form~\eqref{eq:sos-proof} would imply 
\begin{equation}
\Pse{\sum_j h_j^2} = -1 < 0 \, .
\end{equation}

Thus for each degree $d$, there is an SoS proof if and only if there is no pseudoexpectation. These two problems are dual SDPs; a solution to either is a certificate that the other has no solution. In particular, any degree at which the Charlatan can succeed is a lower bound on the degree a refuter needs to prove that no solution exists. In this example, we have shown that degree $4$ is both necessary and sufficient to prove that no three variables in $\{\pm 1\}$ can sum to zero.

\subsection{What does Sum-of-Squares understand?}
\label{sec:sos-understand}

The reader is probably wondering how the SoS framework performs on larger versions of our example. Suppose we have $n$ variables $x_1,\ldots,x_n$. If $n$ is odd, clearly it is impossible to satisfy the system
\begin{align}
\begin{array}{rl}
x_i^2 - 1 &\,=\, 0 \quad \text{for all $i=1,\ldots,n$} \\
\sum_{i=1}^n x_i &\,=\, 0 \, .
\end{array} 
\label{eq:xn-system}
\end{align} 
To put it differently, if you take an odd number of steps in a random walk on the integers, moving one unit to the left or right on each step, there is no way to return to the origin.

It turns out~\cite{grigoriev2001knapsack,grigoriev2001linear,laurent-cut} that any SoS proof of this fact requires degree $n+1$. That is, the Charlatan can construct a pseudoexpectation for polynomials of degree $d$ up to $n-1$. This includes the case $n=3$ we studied above. 

How can the Charlatan do this? Since $x_i^2=1$ for all $i$, it suffices for them to construct pseudoexpectations for the multilinear monomials, i.e., those of the form $x_S = \prod_{i \in S} x_i$ for some set $S \subset \{1,\ldots,n\}$. Furthermore, we can symmetrize over all permutations of the $x_i$, and assume that $\Pse{x_S}$ only depends on their degree $|S|$: semidefinite programming is a convex problem, so symmetric problems have symmetric solutions if any.


Now let $a_k$ denote $\Pse{x_S}$ for $|S|=k$. Equivalently, $a_k = \Pse{x_1 x_2 \cdots x_k}$. We can compute $a_k$ as follows. Suppose I tell you that $n/2$ of the $x_i$ are $+1$, and $n/2$ are $-1$. (Don't ask whether $n/2$ is an integer.) If we choose a uniformly random set of $k$ distinct variables from among the $x_i$, then $a_k$ is the average parity of their product. An enjoyable combinatorial exercise gives, for $k$ even,
\begin{equation}
\label{eq:a-k}
a_k 
= (-1)^{k/2} \,\frac{\binom{n/2}{k/2}}{\binom{n}{k}}
= (-1)^{k/2} 
\frac{(k-1)(k-3)(k-5) \cdots 1}
{(n-1)(n-3)(n-5) \cdots (n-k+1)}
\end{equation}
and $a_k=0$ for $k$ odd. 

Again using the fact that $x_i^2=1$ for all $i$, for any two sets $S,T$ we have $x_S \,x_T = x_{S \triangle T}$ where $\triangle$ denotes the symmetric difference. Thus we define the pseudoexpectation as a bilinear operator that takes monomials $x_S, x_T$ where $|S|,|T| \le d/2$, with matrix elements 
\begin{equation}
\label{eq:laurent-grigoriev}
\bra{x_S} \Psei \ket{x_T} 
= \Pse{x_S \,x_T} 
= \Pse{x_{S \triangle T}}
= a_{|S \triangle T|} \, ,
\end{equation}
which generalizes~\eqref{eq:without-ado} above. As long as $d \le n-1$, it turns out that this $\Psei$ is positive semidefinite~\cite{laurent-cut}; its spectrum can be analyzed using representation theory~\cite{kunisky-moore}. Thus any SoS refutation of the system~\eqref{eq:xn-system} must be of degree at least $d=n+1$. 

This lower bound is tight: any pseudoexpectation on Boolean variables $x_1,\ldots,x_n \in \{\pm 1\}$ of degree $n+1$ must be a true expectation, i.e., must correspond to an actual distribution over the hypercube~\cite{fawzi-saunderson-parrilo}. Thus at degree $n+1$, the Charlatan can no longer produce a convincing pseudoexpectaton unless solutions actually exist. If $n$ is odd, there are no solutions, so by SDP duality there is a refutation of degree $n+1$. 

One way to construct a refutation is as follows. Let $w$ denote $\sum_i x_i$. First we ``prove'' that $w$ is an odd integer between $-n$ and $n$ by finding polynomials $g_1,\ldots,g_n$ such that 
\begin{equation}
\label{eq:odd-proof}
\sum_{i=1}^n g_i(\vec{x}) \,(x_i^2-1) 
= \prod_{t=-n, -n+2,\ldots}^{\ldots,n-2,n} (w-t) 
\, .
\end{equation}
For instance, the reader can check that the three terms inside the square brackets in~\eqref{eq:xyz-proof} sum to  $(w+3)(w+1)(w-1)(w-3)$ where $w=x+y+z$. The polynomials $g_i$ in~\eqref{eq:odd-proof} are guaranteed to exist because, in the ring of polynomials, the set $\{x_i^2-1\}$ spans the set of all polynomials that vanish on $\{\pm 1\}^n$. For the experts, $\{x_i^2-1\}$ is a Gr\"obner basis for this ideal. 

Now we wish to show that some polynomial with $w$ as a factor, say $w^2$, is nonzero. To do this, we find a polynomial $q(w)$ that is everywhere positive and that coincides with $w^2$ at the odd integers between $-n$ and $n$. By polynomial interpolation, we can take $q(w)$ to be even and of degree $n+1$. For $n=3$, for instance, we have
\begin{equation}
q(w) = \frac{1}{8} (w^2-1)^2 + 1 \;\ge\; 1 \, ,
\end{equation}
which we have already written as a sum of squares. 

Since the polynomial $q(w) - w^2$ has these odd integers as roots, it is a multiple of the expression in~\eqref{eq:odd-proof}. 
Putting this together for $n=3$ gives
\begin{equation}
\frac{1}{8} (w+3)(w+1)(w-1)(w+3) \;+\; w^2 = q(w) \, , 
\end{equation}
which is exactly what we wrote in~\eqref{eq:xyz-proof}. 

Now recall that SoS refutations of degree $d$ can be found in polynomial time only if $d$ is a constant. This means that as far as SoS is concerned, proving that~\eqref{eq:xn-system} is unsatisfiable is hard. Clearly SoS doesn't understand parity arguments very well.

Morally, this is because the matrix elements~\eqref{eq:a-k} are analytic functions of $n$: they can't tell whether $n$ is odd or even, or even whether $n$ is an integer or not. To put it differently, binomials like those in the numerator of $a_k$ in~\eqref{eq:a-k} will happily  generalize to half-integer inputs with the help of the Gamma function. After all, there are $\binom{3}{3/2} = 32/(3\pi) = 3.395\dots$ ways to take three steps of a random walk and return to the origin.

The ``hardness'' of this example may make SoS look like a very weak proof system. But parity is a very delicate thing. If $n$ Boolean variables are represented as $\{0,1\}$, then their parity is merely their sum mod $2$; but if we represent them as spins $\pm 1$, the parity is their product, which is of degree $n$. When $n$ is large, we would be amazed to find such a term in the Hamiltonian of a physical system. No observable quantity depends on whether the number of atoms in a block of iron is odd or even. 

The situation seems similar to XORSAT, whose clauses are linear equations mod 2. Its energy landscape has many of the hallmarks of algorithmic hardness, with clusters, frozen variables, and large barriers between solutions~\cite{cocco-etal}. 
See also the discussion in Subsection~\ref{subsec:RSB-and-Clustering} of Section~\ref{sec:OGP}.
In the noise-free case it can be solved in polynomial time using Gaussian elimination over $\Z_2$. But if we add any noise, for instance only requiring that $99\%$ of the XORSAT clauses be satisfied, its algebraic structure falls apart and this algorithmic shortcut disappears. 
So while parity and XORSAT are good cautionary tales, we shouldn't think of them as representative of more generic problems. As we will see next, for many problems with noise, including those involving random matrices and tensors with planted structure, the SoS framework is associated with many algorithms that are conjectured to be optimal.

\subsection{Relaxation and the Sherrington-Kirkpatrick model}

Above we referred to the pseudoexpectation as the work of a charlatan who falsely claims that an unsatisfiable problem has many solutions. But there is another, less adversarial way to describe this character: rather than trying to fool us, they are a \emph{Relaxer} who honestly solves a less-constrained problem, and thus proves bounds on the optimum of the original problem.\footnote{Thanks to Tselil Schramm for suggesting the name ``Relaxer'' for this rehabilitated version of the Charlatan. Perhaps ``Slacker'' would also work in contemporary English.} 

To celebrate the 40th anniversary that inspired this book, let's consider the Sherrington-Kirkpatrick model. Given a coupling matrix $J$  we can write the ground state energy of an Ising spin glass as
\begin{equation}
\label{eq:JX}
E_0
= -\max_{\vec{x} \in \{\pm 1\}^n} \sum_{i < j} J_{ij} x_i x_j 
= -\frac{1}{2} \max_{X \in \mathcal{C}} \tr JX 
\end{equation}
(where we take $J$ to be symmetric and zero on the diagonal). In other words, the energy is quadratic in the spins, but linear in the products $X_{ij} = x_i x_j$. So we just have to maximize a linear function! This is exactly the maximization problem~(\ref{eq:p-spin-ground-state}) when $p=2$, ignoring the $-1/2$ factor.

The tricky part is that we have to maximize $\tr JX$ over a complicated set. In~\eqref{eq:JX}, $\mathcal{C}$ is the set of matrices $X = \ket{x} \bra{x}$ corresponding to actual spin configurations, namely symmetric rank-1 matrices with $\pm 1$ entries and $+1$s on the diagonal. We would get the same maximum if we defined $\mathcal{C}$ to be the polytope of all convex linear combinations of such matrices. But this so-called \emph{cut polytope} has exponentially many facets, making this maximization computationally infeasible~\cite{DL-2009-GeometryCuts}. In the worst case where $J$ is designed by an adversary, it is NP-hard since, for instance, it includes Max Cut as a special case~\cite{Karp-1972-Reducibility}.

We can relax this problem by allowing $X$ to range over some superset $\mathcal{C'}$ of $\mathcal{C}$. Then the maximum of $\tr JX$ will be greater than or equal to the true maximum over $\mathcal{C}$, providing a lower bound on $E_0$. A hopeful goal is to find a set $\mathcal{C'}$ whose structure is simple enough to perform this maximization efficiently, while giving a bound that is not too far from the truth. 

The first attempt we might make is to allow $X$ to range over all positive semidefinite matrices with trace $n$. Call this set $\mathcal{C}_0$:
\begin{equation}
\mathcal{C}_0 = \{ X: X \succeq 0 \text{ and} \tr X = n \} \, . 
\end{equation}
Then
\begin{equation}
\max_{x \in \mathcal{C}_0} \tr JX = n \lambda_{\max}
\end{equation}
where $\lambda_{\max}$ is $J$'s most positive eigenvalue. For the SK model where the $J_{ij}$ are Gaussian with mean $0$ and variance $1/n$, the Wigner semicircle law tells us that, in the limit of large $n$, the spectrum of $J$ is supported on $[-2,2]$. Thus
\begin{equation}
\lim_{n \to \infty} E_0 / n \ge -\frac{\lambda_{\max}}{2} = -1 \, . 
\end{equation}
This is fairly far from  Parisi's solution $E_0/n = -0.7632 $~\cite{parisi79,parisi80}. Can we get a better bound with some other choice of $\mathcal{C}'$?

We can tighten our relaxation by adding any constraint that holds for the true set of matrices $\mathcal{C}$. Let's start with the constraint that $X$'s diagonal entries are $1$. This gives a set of matrices sometimes called the \emph{elliptope}~\cite{laurent-elliptope}, 
\begin{equation}
\mathcal{C}_2 = \{ X : X \succeq 0 \text{ and } X_{ii}=1 \text{ for all $i$} \} \, . 
\end{equation}
We might hope that maximizing $\tr JX$ over $\mathcal{C}_2$ rather than $\mathcal{C}_0$ gives a better bound on the energy. Unfortunately, this is not the case: for any constant $\eps > 0$, with high probability there is an $X \in \mathcal{C}_2$ such that $\tr JX \ge 2-\eps$. We will sketch the proof of~\cite{montanari-sen}. 

First let $v_\lambda$ denote the eigenvector of $J$ with eigenvalue $\lambda$, normalized so that $|v_\lambda|^2=1$. Let $m$ denote the number of eigenvalues in the interval $[2-\eps,2]$. These eigevalues span a low-energy subspace where $E_0 \approx -1$. Now define $Y$ as 
\begin{equation}
Y = \frac{n}{m} \sum_{\lambda \in [2-\eps,2]} \ket{v_\lambda}\bra{v_\lambda} \, .
\end{equation}
That is, $Y$ is $n/m$ times the projection operator onto this subspace. Thus $Y \succeq 0$ and $\tr JY \ge (2-\eps)n$. 

We can write $Y$'s diagonal entries as
\begin{equation}
Y_{ii} = \frac{n}{m} 
\sum_\lambda (v_\lambda)_i^2 \, .
\end{equation}
Since the distribution of Gaussian random matrices is rotationally invariant, the $v_\lambda$ are distributed as a uniformly random set of $m$ orthonormal vectors in $n$ dimensions. Thus the $(v_\lambda)_i^2$ are asymptotically independent, and are $1/n$ on average. As a result, each $Y_{ii}$ is concentrated around $1$.

To turn $Y$ into an $X$ such that $X_{ii}=1$ holds exactly, define $D$ as the diagonal matrix $D_{ii}=Y_{ii}$ and let 
\begin{equation}
\label{eq:montanari-sen}
X = D^{-1/2} Y D^{-1/2} \, . 
\end{equation}
Clearly $X \succeq 0$. Moreover, since $D$ itself is close to the identity, we have $\tr JX = \tr JY$ up to a vanishing error term. Since $X \in \mathcal{C}_2$, we have shown that $\mathcal{C}_2$ doesn't give a bound any better than the simple spectral bound provided by $\mathcal{C}_0$.

The alert reader will note that $\mathcal{C}_2$ is exactly the set of pseudoexpectations $\Psei$ that a degree-2 charlatan can choose from. If $X_{ij} = \Pse{x_i x_j}$, then $X \succeq 0$ and $X_{ii} = \Pse{x_i^2} = 1$. So whether we regard $X$ as the solution to a relaxed problem or a false claim about the covariances $\Exp{x_i x_j}$, we have shown that degree-2 SoS proofs cannot establish a bound better than $E_0/n > -1$ on the SK ground state energy. That is, they are incapable of refuting the claim that there are states with energy  $-1+\eps$ or below, for arbitrarily small $\eps$. 

(There is a subtlety here. The refuter's goal is not to understand the typical ground state energy of the SK model, but to provide ironclad proofs for individual realizations $J$ that their ground state energy is above a certain point. What we have shown is that, for most realizations $J$, there is no degree-2 proof that its ground state energy is noticeably above $-1$.)

We should also note that, just as $\mathcal{C}$ is the set of matrices $X = \ket{x} \bra{x}$ where the $x_i = \pm 1$ are Ising spins, $\mathcal{C}_2$ is the set of matrices $X = \ket{x} \bra{x}$ where the $x_i$ are $n$-dimensional vectors with $|x_i|^2 = 1$. So while Ising spins can't achieve the covariances $X_{ij} = \Pse{x_i x_j}$ that the Charlatan claims, these vector-valued spins can achieve them in the sense that $X_{ij} = \inner{x_i}{x_j}$. 

This is the heart of the  Goemans-Williamson approximation algorithm for Max Cut \cite{goemans-williamson}---or, in physics terms, bounding the ground-state energy of an antiferromagnet. In Max Cut, our goal is to assign a spin $x_i = \pm 1$ to each vertex, and maximize the number $w$ of edges whose spins are opposite. For a graph with $m$ edges and adjancency matrix $A$, this is
\begin{equation}
w = \frac{1}{2}
\big( m - \bra{x} \!A\! \ket{x} \big) \, .
\end{equation}
If we relax this problem by letting the $x_i$ be unit-length vectors in $\R^n$ instead of just $\pm 1$, this becomes an SDP that we can solve in polynomial time. It can be shown that this relaxation increases $w$ by a factor of at most $1/0.878=1.138...$, so the optimum of this relaxation is not too far from that of the original problem. 

We do not know whether going to higher-degree SoS improves this approximation ratio. If we assume the Unique Games Conjecture (a plausible strengthening of $\textrm{P} \ne \textrm{NP}$) then no polynomial-time algorithm can do better than Goemans-Williamson~\cite{khot-etal-unique-games-maxcut}.\footnote{This is usually presented the other way around. If we round the relaxed solution to $\pm 1$ spins by cutting $\R^n$ with a random hyperplane, the Goemans-Williamson algorithm gives a cut that is at least $0.878$ times the optimum, and the Unique Games Conjecture implies that this cannot be improved. The same argument~\cite{khot-etal-unique-games-maxcut} implies an upper bound on the relaxed solution. (Thanks to Tim Kunisky for pointing this out).} This suggests that going to degree 4, 6, and so on doesn't give a better algorithm, but even for degree 4 this is an open question. 

On the other hand, for the SK model it was recently shown~\cite{ghosh-etal-sk} that higher-degree SoS does not improve our bounds on the ground state energy, as we will see next.

\subsection{Beyond degree 2}

Can SoS proofs of some constant degree $d > 2$ prove a tighter bound on the ground state energy $E_0$ of the Sherrington-Kirkpatrick model? Do higher-degree polynomials help us go beyond the simple spectral bound $E_0 \ge -1$?

The Charlatan's job for $d=4$ is already quite interesting. In addition to providing $X \in \mathcal{C}_2$, they now have to provide an $\binom{n}{2}$-dimensional matrix $X^{(4)}$, with rows and columns for each pair $(i,j)$, such that 
\begin{equation}
X^{(4)}_{(i,j),(k,\ell)} = \Pse{x_i x_j x_k x_\ell} \, . 
\end{equation}
Thus $X^{(4)}$ must have the symmetries of a symmetric four-index tensor,
\begin{equation}
\label{eq:x4-1}
X^{(4)}_{(i,j),(k,\ell)} 
= X^{(4)}_{(i,k),(j,\ell)} 
= X^{(4)}_{(i,\ell),(j,k)} \, .
\end{equation}
In addition, $X^{(4)}$ needs to be consistent with the degree-2 pseudexpectations and the constraint $x_i^2=1$. Thus
\begin{gather}
X^{(4)}_{(i,j),(i,k)} 
= \Pse{x_i^2 x_j x_k} 
= \Pse{x_j x_k} 
= X_{jk} 
\label{eq:x4-2} \\
X^{(4)}_{(i,j),(i,j)}
= \Pse{x_i^2 x_j^2} 
= 1 \, . 
\label{eq:x4-3}
\end{gather}
(We saw these relations in Section~\ref{sec:sos-understand} where we wrote $x_S x_T = x_{S \triangle T}$.) Finally, as always $X^{(4)}$ must be positive semidefinite,
\begin{equation}
\label{eq:x4-4}
X^{(4)} \succeq 0 \, .
\end{equation}

The energy $E=-(1/2) \tr JX$ is still a function of the second-order pseudoexpectation $X$. But not all matrices $X$ in $\mathcal{C}_2$ can be extended to fourth order in this way: the set
\begin{equation}
\mathcal{C}_4 
= \{ X \in \mathcal{C}_2 : 
\text{$\exists X^{(4)}$
such that~\eqref{eq:x4-1}--\eqref{eq:x4-4} holds} \}
\end{equation}
is a proper subset of the elliptope $\mathcal{C}_2$. In other words, armed with degree-4 SoS proofs, a refuter can prove some new constraints on the covariances $X_{ij} = x_i x_j$ that go beyond $X_{ii}=1$ and $X \succeq 0$.

For example, consider any three Ising spins, $x_i$, $x_j$, and $x_k$. Their products $(x_i x_j, x_j x_k, x_i x_k)$ can only take the values $(1,1,1)$, $(1,-1,-1)$, $(-1,1,-1)$, and $(-1,-1,1)$. Thus the expectation of their products $(X_{ij}, X_{jk}, X_{ik})$ must lie in the convex hull of these four vectors, namely the tetrahedron with these four vertices. The facets of this tetrahedron are the linear inequalities
\begin{align}
X_{ij} + X_{jk} + X_{ik} + 1 
& \ge 0 \label{eq:tet1} \\
X_{ij} - X_{jk} - X_{ik} + 1 
& \ge 0 \label{eq:tet2} \\
-X_{ij} + X_{jk} - X_{ik} + 1 
& \ge 0 \label{eq:tet3} \\
-X_{ij} - X_{jk} + X_{ik} + 1 
& \ge 0 \label{eq:tet4} \, .
\end{align}
We have already seen a pseudoexpectation in $\mathcal{C}_2$ that violates the first of these inequalities---namely~\eqref{eq:without-ado} where $X_{ij} = X_{jk} = X_{ik} = -1/2$. Thus we cannot prove these inequalities with degree-2 sum-of-squares. But we can prove them with degree 4, and we already have! After all, we can rewrite~\eqref{eq:tet1} as
\begin{equation}
\Pse{x_i x_j + x_j x_k + x_i x_k + 1} \ge 0 \, . 
\end{equation}
But if $x_i^2=x_j^2=x_k^2=1$ this is equivalent to 
\begin{equation}
\Pse{ (x_i + x_j + x_k)^2 } \ge 1 \, . 
\end{equation}
Looking again at our proof~\eqref{eq:xyz-proof} that no three spins can sum to zero, the reader will see that we in fact proved that $(x+y+z)^2 \ge 1$ whenever $x^2=y^2=z^2=1$. The symmetry operations $x \mapsto -x$, $y \mapsto -y$, and $z \mapsto -z$ give similar proofs of~\eqref{eq:tet2}--\eqref{eq:tet4}. 

Thus any matrix $X$ that violates these ``triangle inequalities'' can be refuted by degree-4 sum-of-squares. More generally, since any $t+1$ pseudoexpectation on $t$ spin variables is a true expectation~\cite{fawzi-saunderson-parrilo}, any linear inequality on the covariances of $t$ spins---or equivalently any inequality that involves a $t \times t$ principal minor of $X$---can be proved with degree $t+1$ sum-of-squares.

Perhaps these and other degree-4 constraints will finally give a better bound on $E_0$? Sadly---or happily if you love computational hardness---they do not. In fact, no constant degree can refute the claim that some spin confirugation lies in the low-energy subspace, and thus prove a bound tighter than $\tr JX \le 2$ or $E_0 \ge -1$. 

One intuition for this is that for natural degree-2 pseudoexpectations, like the $X$ we constructed above~\eqref{eq:montanari-sen} by projecting onto the low-energy subspace, triangle inequalities and their generalizations already hold with room to spare. In the SK model we typically have $\Pse{x_i x_j} = O(1/\sqrt{n})$, so \eqref{eq:tet1}--\eqref{eq:tet4} all read $1+O(1/\sqrt{n}) \ge 0$. Thus, with perhaps a slight perturbation to make it positive definite and full rank, $X$ is already deep inside the elliptope $\mathcal{C}_2$, and is not refuted by the additional inequalities we can prove with low-degree SoS proofs. 

There are several ways to make this intuition rigorous. One is to explicitly construct higher-degree pseudoexpectations $X^{(4)}$, $X^{(6)}$, and so on that extend $X$ in a natural way, somewhat like a cluster expansion in physics. For instance, we could define
\begin{equation}
\label{eq:x4-tim}
X^{(4)}_{(i,j),(k,\ell)}
= X_{ij} X_{k\ell} 
+ X_{ik} X_{j\ell}
+ X_{i\ell} X_{jk} 
- 2 \sum_{m=1}^n X_{im} X_{jm} X_{km} X_{\ell m} \, . 
\end{equation}
This expression has the permutation symmetry of~\eqref{eq:x4-1}. The first three terms look like Wick's theorem or Isserlis' theorem for the moments of Gaussian variables~\cite{isserlis}; the reader can check that by cancelling two of these terms when $k=\ell$, the sum over $m$ ensures the consistency relations~\eqref{eq:x4-2} and \eqref{eq:x4-3} to leading order. A small perturbation then satisfies these conditions exactly~\cite{kunisky-bandeira-deg4} and it is relatively easy to show that the result is positive semidefinite; see also~\cite{mohanty-raghavendra-xu}. A similar approach works for degree 6~\cite{kunisky-deg6}.

\subsection{Pseudocalibration and clever planted models}

While constructions  like~\eqref{eq:x4-tim} could probably be carried out for higher degree, the recent proof~\cite{ghosh-etal-sk} that no constant degree of SoS can improve the bound on $E_0$ comes from a different direction called \emph{pseudocalibration}~\cite{barak-etal-psuedocalibration,raghavendra2018high}. 

In pseudocalibration, the Charlatan claims that the data is generated by a planted model where the claimed solution is built in, rather than the (true) null model. In the Sherrington-Kirkpatrick model this means pretending that the couplings $J$ have been chosen so that some Boolean vector $x \in \{\pm 1\}^n$ achieves the spectral bound $E_0 = -1$. 

If we can construct a pseudoexpectation around this idea, then low-degree SoS can't tell the difference between the null model and the planted model. In particular, it can't prove that the planted solution doesn't exist.

Following~\cite{raghavendra2018high}, we can briefly describe pseudocalibration as follows. We consider two joint distributions on a signal $x$ and observed data $Y$. In both cases, we choose $x$ from a prior $P(x)$. In the null model, we choose $Y$ independently of $x$ with probability $\Pnull(Y)$; in the planted model, we choose $Y$ with probability $\Pplanted(Y \mid x)$. Thus 
\begin{gather*}
\Pnull(x,Y) = \Pnull(Y) \,P(x) \\
\Pplanted(x,Y) 
= \Pplanted(Y \mid x) \,P(x) 
= \Pplanted(Y) \,\Pplanted(x \mid Y) \, , 
\end{gather*}
where $\Pplanted(Y) = \mathbb{E}_{x \sim P(x)} \Pplanted(Y \mid x)$ is $Y$'s likelihood in the planted model.

In the Charlatan's first attempt, they define the  pseudoexpectation of a function $q(x)$ as its true expectation given $Y$, but reweighted to change the null model into the planted one:
\begin{align}
\Pse{q(x) \mid Y} 
&= \mathbb{E}_{x \sim P(x)} \left[ \frac{\Pplanted(x,Y)}{\Pnull(x,Y)} \,q(x) \right] 
\nonumber \\
&= \mathbb{E}_{x \sim P(x)} \left[ \frac{\Pplanted(Y) \,P_1(x \mid Y)}{\Pnull(Y) \,P(x)} \,q(x) \right] \nonumber \\
&= \frac{\Pplanted(Y)}{\Pnull(Y)} \,\mathbb{E}_{x \sim P_1(x \mid Y)} q(x) \, . 
\label{eq:first-attempt}
\end{align}
That is, the pseudoexpectation of $q(x)$ is its true expectation in the  posterior distribution $P_1(x \mid Y)$, multiplied by the likelihood ratio $P_1(Y)/P_0(Y)$. 

This pseudoexpectation is proportional to a true expectation, albeit over another distribution. Thus it is positive semidefinite, $\Pse{q^2} \ge 0$. Similarly, if $P(x)$ and therefore $P(x \mid Y)$ are supported on $x$ satisfying some constraint $f_i(x)=0$, then $\Pse{f_i q}=0$ for any $q$. 

Moreover, \eqref{eq:first-attempt} gives any function of $x$ and $Y$ the expectation over the null model that it would have in the planted model, 
\begin{equation}
\label{eq:looks-like-planted}
\mathbb{E}_{Y \sim \Pnull} \Pse{q(x,Y)} 
= \mathbb{E}_{(x,Y) \sim P_0} 
\left[ \frac{\Pplanted(x,Y)}{\Pnull(x,Y)} \,q(x,Y) \right]
= \mathbb{E}_{(x,Y) \sim P_1} q(x,Y)
\, .
\end{equation}
where we took the average over $Y$ as well as $x$.

On the other hand, for individual $Y$ we have some trouble. For instance, \eqref{eq:first-attempt} gives $\Pse{1 \mid Y}=P_1(Y)/P_0(Y)$, the likelihood ratio instead of $1$. This would make it easy to catch the Charlatan whenever the null and planted models can be distinguished information-theoretically. Moreover, while the planted model guarantees that $Y$ has a solution $x$, most $Y$ drawn from the null model have no such solution. In that case we have $P_1(Y)=0$, and the posterior distribution $P_1(x \mid Y)$ is undefined.

We can fix both these problems by projecting 
$\Pse{q(x) \mid Y}$ 
into the space of low-degree polynomials, both in $x$ and in $Y$. In other words, we take its Taylor series in $x$ and $Y$ up to some degree. For Boolean variables, this is equivalent to keeping just the low-frequency part of the Fourier spectrum; in some cases, we might project onto a suitable set of orthogonal polynomials. This preserves the appearance~\eqref{eq:looks-like-planted} of of the planted model for functions of low degree in $x$ and $Y$.

If all goes well, this projection smooths the likelihood ratio, keeping it concentrated around its expectation $1$. It also smooths the posterior distribution $P_1(x \mid Y)$ as a function of $Y$, extending it from the small set of $Y$ produced by the planted model (for instance, the few instances of the SK model where $E_0=-1$) to the more generic $Y$ produced by the null model. 

However, the Charlatan has to preserve enough of the dependence on $Y$ to make $\Pse{q \mid Y}$ convincing. To do this for $q(x)$ of degree $d$, they typically need to preserve terms in $Y$ up to some sufficient degree $D > d$.

Showing that $\Psei$ remains positive semidefinite after this projection, and that it continues to satisfy the constraints $\Pse{f_i}=0$, can involve summing over many combinatorial terms. This was first done for the Planted Clique problem~\cite{barak-etal-psuedocalibration}. While each application since then has involved special-purpose calculations, several conjectures~\cite{raghavendra2018high} offer general principles by which this program might be extended.

The projection of $\Pse{1 \mid Y} = P_1(Y)/P_0(Y)$ into low-degree polynomials in $Y$ is of its own interest: it is the \emph{low-degree likelihood ratio}. If it is usually close to $1$ in the null model but is large in the planted model, then it provides a polynomial-time hypothesis test for distinguishing between these two. Thus showing that it has bounded variance in the null model is in itself evidence of computational hardness~\cite{hopkins2018statistical}. In particular, \cite{bandeira-kunisky-wein-pca} showed that the degree-$D$ likelihood ratio fails to improve the bound on the SK model for any $D=o(n/\log n)$. This does not in itself prove that SoS fails up to this degree, but the two approaches are closely related.

We conclude this section by discussing the choice of planted model. Proving that refutation is hard might require a clever way to hide a solution, as opposed to the standard spiked matrices and tensors. For instance, to prove their SoS lower bounds on the Sherrington-Kirkpatrick model, \cite{ghosh-etal-sk} related a planted model proposed by~\cite{mohanty-raghavendra-xu} where a random subspace (i.e., the low-energy subspace) contains a Boolean vector to a model of Gaussian random vectors, where in the planted case these vectors belong to two parallel hyperplanes.

More generally, there is a long history in physics and computer science of ``quiet'' planting, in order to make the solution as difficult as possible to detect~\cite{krzakala2009hiding,zdeborova2011quiet}.  The quieter the planting, the harder it is to distinguish from the null model. In this case, we want the planting to be \emph{computationally} quiet~\cite{bandeira-kunisky-wein-pca}, and in particular to match the low-degree moments of the null distribution. For instance, rather than the usual spiked model where we add a rank-1 perturbation to a Gaussian random matrix $J$---which disturbs the entire spectrum---we can plant a large eigenvalue more quietly by  increasing the eigenvalue of a specific eigenvector~\cite{bandeira-etal-cuts-colorings}.

\subsection{Optimal algorithms and the curious case of tensor PCA}

We've talked a lot about what SoS algorithms can't do. But for many problems they seem to be optimal, performing as well as any polynomial-time algorithm can. For Max Cut and the Sherrington-Kirkpatrick model, we've seen evidence that this is the case even at degree 2. 

Thus in many cases, SoS algorithms seem to succeed or fail at the same place where physics suggests a hard/easy transition. Even when these thresholds don't coincide exactly, they often have the same scaling and thus differ by a constant. For example, degree-2 SoS---also known as the Lov\'asz $\vartheta$ function---can refute graph colorings in random regular graphs within a factor of $4$ of the Kesten-Stigum transition~\cite{banks-kleinberg-moore}, and it's possible that higher-degree SoS does better. 

While refuting the existence of a planted solution lets SoS solve the detection problem---distinguishing the null from a planted model---a refinement of this idea often yields algorithms for reconstruction as well. Roughly speaking, if we can refute the existence of a solution when it doesn't exist, we can often find it when it does~\cite{raghavendra2018high}. 

To see how this works, consider a planted model, and let $x^*$ denote the ground truth. Let $\phi(x)$ be some polynomial for which $\phi(x^*) \le \phi^*$: for instance, in PCA, $\phi(x)$ could be the $\ell_2$ distance between the signal matrix $\ket{x} \bra{x}$ and the observed matrix $Y$. Now suppose there is a degree-$d$ refutation of the claim that there are any good solutions far from the ground truth: that is, a proof that if $\phi(x) \le \phi^*$ then $|x-x^*|^2 \le \eps$. Then any degree-$d$ pseudoexpectation must claim that $|\Pse{x}-x^*|^2 \le \eps$, and $\Pse{x}$ is a good estimate of $x^*$. 

This approach yields efficient algorithms for many problems~\cite{barak-kelner-steurer-rounding,barak-moitra-matrix-completion}, including tensor PCA~\cite{hopkins-shi-steurer-tensor-pca}. But for tensor PCA in particular, a curious gap appeared between algorithms and physics. Recall from Section~\ref{sec:spiked-tensor} that tensor PCA, a.k.a.\ the spiked tensor model, is a planted model of $p$-index tensors defined by 
\begin{equation}
Y = \lambda u^{\otimes p} + J \, .
\end{equation}
Here $\lambda$ is the signal-to-noise ratio, the planted vector $u$ is normalized so that $|u|^2=n$, and the noise tensor $J$ is permutation-symmetric with Gaussian entries  $\mathcal{N}(0,1)$. 
The information-theoretic transition occurs at $\lambda = \lambda_c  n^{-(p-1)/2}$ for a constant $\lambda_c$ depending on $p$ and $u$'s prior~\cite{richard2014statistical,lesieur2017statistical}.

The best known polynomial-time algorithms, on the other hand, require a considerably larger signal-to-noise ratio, $\lambda \gtrsim n^{-p/4}$. One such algorithm, called ``tensor unfolding,'' reinterprets $Y$ as a matrix and iteratively applies PCA to it. For $p=4$, for instance, we treat $Y$ as an $n^2 \times n^2$ matrix $Y_{ij,k\ell}$ and find its leading eigenvector $v$. Since $v \approx u \otimes u$, we then treat $v$ as an $n \times n$ matrix and estimate $u$ as its leading eigenvector. At each stage we unfold the tensor into a matrix which is as square as possible.

Other algorithms, that also succeed for $\lambda \gtrsim n^{-p/4}$, can be derived directly from sum-of-squares~\cite{hopkins-etal-fast-spectral}. Conversely, SoS lower bounds suggest that there is no polynomial-time algorithm if $\lambda \lesssim n^{-p/4}$, so this appears to be the algorithmic threshold~\cite{hopkins2017power}.\footnote{Our notation $\gtrsim$ and $\lesssim$ suppresses logarithmic factors. These are consequences of matrix Chernoff bounds, and could probably be removed.}

On the other hand, physics-based algorithms such as belief propagation and its asymptotic cousin approximate message passing (AMP), as well as Langevin dynamics, all fail unless $\lambda \gtrsim n^{-1/2}$, making these algorithms suboptimal whenever $p \ge 3$~\cite{richard2014statistical,anandkumer-tensor-pca}. This does not contradict conjectures of optimality from section \ref{sec:is_it_hard} as those were restricted to the scaling of parameters corresponding to the information-theoretical regime which in this case is $\lambda \approx  n^{-(p-1)/2}$. Never-the-less, focusing on the regime discussed here, does sum-of-squares know something that physics doesn't?

This conundrum has a satisfying answer~\cite{wein-alaoui-moore}: in the scaling regime $\lambda \gtrsim n^{-(p-1)/2}$ we were using the wrong physics. Belief propagation keeps track of pairwise correlations. When we compute the Bethe free energy, we pretend that the Gibbs distribution, i.e., the posterior distribution $P(x \mid Y)$, has the form
\begin{equation}
P(x) = \prod_i \mu_i(x_i) \times \prod_{(i,j)} 
\frac{\mu_{ij}(x_i,x_j)}{\mu_i(x_i) \,\mu_j(x_j)} 
\end{equation}
where $\mu_i$ and $\mu_{ij}$ are one- and two-point marginals. Minimizing the resulting free energy is equivalent to finding fixed points of belief propagation~\cite{YedidiaFreemanWeiss}.

But when $p \ge 3$, it becomes vital to consider correlations between clusters of $p$ variables. This gives rise to a hierarchy of free energies due to~\cite{kikuchi}. For $p=3$, for instance, we assume that the Gibbs distribution has the form
\begin{equation}
P(x) 
= \prod_i \mu_i
\times \prod_{(i,j)} 
\frac{\mu_{ij}}{\mu_i \,\mu_j} 
\times \prod_{(i,j,k)}
\frac{\mu_{ijk} \,\mu_i \,\mu_j \,\mu_k}
{\mu_{ij} 
\,\mu_{jk}
\,\mu_{ik}
}
\end{equation}
(where for readability we suppress $(x_i)$, $(x_i,x_j)$, and so on). At each level of this approximation, we correct for overcounting smaller clusters. Taking the logarithm of this expression and averaging over $x$ gives an inclusion-exclusion-like formula for the entropy. 

There are several ways one might turn this into a spectral algorithm. One is to write an iterative algorithm to minimize the free energy. This gives rise to a generalization of belief propagation in which each variable sends messages to clusters of up to $p-1$ variables with which it interacts ~\cite{yedidia-kikuchi-nips,YedidiaFreemanWeiss-kikuchi}. One could then linearize this message-passing algorithm around a trivial fixed point, producing a operator analogous to the non-backtracking operator for belief propagation~\cite{spectral-redemption,bordenave-lelarge-massoulie}. 

An alternate approach is to compute the Hessian of the free energy at a trivial fixed point, generalizing the use of the Bethe Hessian for spectral clustering in graphs~\cite{saade-bethe-hessian}. This gives rise to the following operator. For a set $U=\{s_1,\ldots,s_p\}$ with $|U|=p$, let $Y_U$ denote $Y_{s_1,\ldots,s_p}$. Fix $\ell \ge p/2$. Then define the following $\binom{n}{\ell}$-dimensional operator, whose rows and columns are indexed by sets $S,T$ with $|S|=|T|=\ell$:
\begin{equation}
M_{S,T} = \begin{cases}
Y_{S \triangle T} 
& \text{if $|S \triangle T|=p$} \\
0 & \text{otherwise} \, , 
\end{cases}
\end{equation}
where $\triangle$ again denotes the symmetric difference. 

The spectral norm of $M$ can be used as a test statistic to distinguish the planted model from the null model where $\lambda=0$. In addition, the leading eigenvector of $M$ points approximately to the minimum of the free energy, and a voting procedure yields a good estimate of the signal $u$. This yields polynomial-time algorithms for detection and reconstruction whenever $\lambda \gtrsim n^{-p/4}$, matching the SoS threshold. Thus the marriage of algorithms and statistical physics is redeemed~\cite{wein-alaoui-moore}.

The same analysis matches a continuum  of subexponential-time algorithms at smaller values of $\lambda$~\cite{bhattiprolu_et_al} and yields a simpler refutation of random constraint satisfaction problems at high clause densities~\cite{raghavendra-rao-schramm}. These ``Kikuchi matrices'' have additional applications, e.g.~\cite{guruswami-kothari-manohar}.

\section{Conclusion}

What does the future hold? As our understanding of algorithms deepens, we hope to understand the universal characteristics that make problems easy or hard, unifying larger and larger classes of polynomial-time algorithms and connecting them rigorously with physical properties of the energy landscape. Very recently, 
\cite{bandeira-etal-franz-parisi} connected the low-degree likelihood ratio with the Franz-Parisi potential, adding to the evidence that free energy barriers imply computational hardness. We will know much more in a few years than we know now.

\section*{Acknowledgments}

We are deeply grateful to Tim Kunisky, Tselil Schramm, and Alex Wein for helpful comments on Section~\ref{sec:sos}. We also thank Freya Behrens, Giovanni Piccioli, Paula Mürmann, Yatin Dandi, Emanuele Troiani for their useful comments on the manuscript. C.M. is supported by NSF grant BIGDATA-1838251, D.G. acknowledged the funding from grant DMS-2015517.

\section*{Bibliography}


\bibliographystyle{iopart-num}    %
\bibliography{cris,bibsample,bibliography-2}      

\providecommand{\newblock}{}
\begin{thebibliography}{100}
\expandafter\ifx\csname url\endcsname\relax
  \def\url#1{{\tt #1}}\fi
\expandafter\ifx\csname urlprefix\endcsname\relax\def\urlprefix{URL }\fi
\providecommand{\eprint}[2][]{\url{#2}}

\bibitem{nature_of_computation}
Moore C and Mertens S 2011 {\em The Nature of Computation\/} (Oxford University
  Press)

\bibitem{cook1971complexity}
Cook S~A 1971 The complexity of theorem-proving procedures {\em Proceedings of
  the 3rd Annual Symposium on Theory of computing\/} pp 151--158

\bibitem{fu1986application}
Fu Y and Anderson P~W 1986 {\em Journal of Physics A: Mathematical and
  General\/} {\bf 19} 1605

\bibitem{cheeseman1991really}
Cheeseman P~C, Kanefsky B, Taylor W~M {\em et~al.\/} 1991 Where the really hard
  problems are. {\em Ijcai\/} vol~91 pp 331--337

\bibitem{monasson1999determining}
Monasson R, Zecchina R, Kirkpatrick S, Selman B and Troyansky L 1999 {\em
  Nature\/} {\bf 400} 133--137

\bibitem{donoho2018optimal}
Donoho D~L, Gavish M and Johnstone I~M 2018 {\em Annals of Statistics\/} {\bf
  46} 1742

\bibitem{lesieur2017constrained}
Lesieur T, Krzakala F and Zdeborov{\'a} L 2017 {\em Journal of Statistical
  Mechanics: Theory and Experiment\/} {\bf 2017} 073403

\bibitem{sherrington1975solvable}
Sherrington D and Kirkpatrick S 1975 {\em Physical Review Letters\/} {\bf 35}
  1792

\bibitem{babacan2012sparse}
Babacan S~D, Luessi M, Molina R and Katsaggelos A~K 2012 {\em IEEE Transactions
  on Signal Processing\/} {\bf 60} 3964--3977

\bibitem{moore_eatcs}
Moore C 2017 {\em Bull. {EATCS}\/} {\bf 121}
  \urlprefix\url{http://eatcs.org/beatcs/index.php/beatcs/article/view/480}

\bibitem{gardner1988optimal}
Gardner E and Derrida B 1988 {\em Journal of Physics A: Mathematical and
  general\/} {\bf 21} 271

\bibitem{krauth1989storage}
Krauth W and M{\'e}zard M 1989 {\em Journal de Physique\/} {\bf 50} 3057--3066

\bibitem{lecun2015deep}
LeCun Y, Bengio Y and Hinton G 2015 {\em Nature\/} {\bf 521} 436--444

\bibitem{achlioptas2006solution}
Achlioptas D and Ricci-Tersenghi F 2006 On the solution-space geometry of
  random constraint satisfaction problems {\em Proceedings of the 38th Annual
  Symposium on Theory of Computing\/} pp 130--139

\bibitem{mezard2005clustering}
M{\'e}zard M, Mora T and Zecchina R 2005 {\em Physical Review Letters\/} {\bf
  94} 197205

\bibitem{gamarnik2021overlap}
Gamarnik D 2021 {\em Proceedings of the National Academy of Sciences\/} {\bf
  118}

\bibitem{o2014analysis}
O'Donnell R 2014 {\em Analysis of {B}oolean {F}unctions\/} (Cambridge
  University Press)

\bibitem{gamarnik2021overlapAukosh}
Gamarnik D and Jagannath A 2021 {\em The Annals of Probability\/} {\bf 49}
  180--205

\bibitem{gamarnik2020lowFOCS}
Gamarnik D, Jagannath A and Wein A~S 2020 Low-degree hardness of random
  optimization problems {\em 61st Annual Symposium on Foundations of Computer
  Science\/}

\bibitem{wein2020optimal}
Wein A~S 2020 {\em Mathematical Statistics and Learning. To appear\/}

\bibitem{gamarnik2020hardness}
Gamarnik D, Jagannath A and Wein A~S 2020 {\em arXiv preprint
  arXiv:2004.12063\/}

\bibitem{farhi2020quantumRandom}
Farhi E, Gamarnik D and Gutmann S 2020 {\em arXiv preprint arXiv:2004.09002\/}

\bibitem{chou2021limitations}
Chou C~N, Love P~J, Sandhu J~S and Shi J 2021 {\em arXiv preprint
  arXiv:2108.06049\/}

\bibitem{basso2022performance}
Basso J, Gamarnik D, Mei S and Zhou L 2022 {\em arXiv preprint
  arXiv:2204.10306\/}

\bibitem{parisi1980sequence}
Parisi G 1980 {\em Journal of Physics A: Mathematical and General\/} {\bf 13}
  L115

\bibitem{MezardParisiVirasoro}
Mézard M, Parisi G and Virasoro M~A 1987 {\em Spin-Glass Theory and Beyond,
  {\rm {V}ol 9 of } {L}ecture {N}otes in {P}hysics\/} (World Scientific,
  Singapore)

\bibitem{GuerraTon}
FGuerra and FLToninelli 2002 {\em Commun. Math. Phys.\/} {\bf 230} 71--79

\bibitem{talagrand2006parisi}
Talagrand M 2006 {\em Annals of Mathematics\/} {\bf 163} 221--263

\bibitem{panchenko2013parisi}
Panchenko D 2013 {\em Annals of Mathematics\/} {\bf 177} 383--393

\bibitem{panchenko2013sherrington}
Panchenko D 2013 {\em The {S}herrington-{K}irkpatrick model\/} (Springer
  Science \& Business Media)

\bibitem{crisanti1992sphericalp}
Crisanti A and Sommers H~J 1992 {\em Zeitschrift f{\"u}r Physik B Condensed
  Matter\/} {\bf 87} 341--354

\bibitem{subag2021following}
Subag E 2021 {\em Communications on Pure and Applied Mathematics\/} {\bf 74}
  1021--1044

\bibitem{montanari2021optimization}
Montanari A 2021 {\em SIAM Journal on Computing\/}  FOCS19--1

\bibitem{el2021optimization}
El~Alaoui A, Montanari A and Sellke M 2021 {\em The Annals of Probability\/}
  {\bf 49} 2922--2960

\bibitem{chen2019suboptimality}
Chen W~K, Gamarnik D, Panchenko D and Rahman M 2019 {\em The Annals of
  Probability\/} {\bf 47} 1587--1618

\bibitem{auffinger2018energy}
Auffinger A and Chen W~K 2018 {\em Advances in Mathematics\/} {\bf 330}
  553--588

\bibitem{chatterjee2009disorder}
Chatterjee S 2009 {\em arXiv preprint arXiv:0907.3381\/}

\bibitem{chen2018disorder}
Chen W~K, Panchenko D {\em et~al.\/} 2018 {\em The Annals of Applied
  Probability\/} {\bf 28} 1356--1378

\bibitem{gamarnik2014limits}
Gamarnik D and Sudan M 2017 {\em Annals of Probability\/} {\bf 45} 2353--2376

\bibitem{huang2021tight}
Huang B and Sellke M 2021 {\em arXiv preprint arXiv:2110.07847\/}

\bibitem{bresler2021algorithmicFOCS2021}
Bresler G and Huang B 2021 {\em FOCS 2021\/}

\bibitem{rossman2008constant}
Rossman B 2008 On the constant-depth complexity of k-clique {\em Proceedings of
  the fortieth annual ACM symposium on Theory of computing\/} pp 721--730

\bibitem{rossman2010average}
Rossman B 2010 {\em Average-case complexity of detecting cliques\/} Ph.D.
  thesis Massachusetts Institute of Technology

\bibitem{li2017ac}
Li Y, Razborov A and Rossman B 2017 {\em SIAM Journal on Computing\/} {\bf 46}
  936--971

\bibitem{rossman2018lower}
Rossman B 2018 Lower bounds for subgraph isomorphism {\em Proceedings of the
  International Congress of Mathematicians: Rio de Janeiro 2018\/} (World
  Scientific) pp 3425--3446

\bibitem{AchlioptasCojaOghlanRicciTersenghi}
Achlioptas D, Coja-Oghlan A and Ricci-Tersenghi F 2011 {\em Random Structures
  and Algorithms\/} {\bf 38} 251--268

\bibitem{aubin2019storage}
Aubin B, Perkins W and Zdeborov{\'a} L 2019 {\em Journal of Physics A:
  Mathematical and Theoretical\/} {\bf 52} 294003

\bibitem{abbe2021proof}
Abbe E, Li S and Sly A 2021 {\em arXiv preprint arXiv:2102.13069\/}

\bibitem{abbe2021binary}
Abbe E, Li S and Sly A 2021 {\em arXiv preprint arXiv:2111.03084\/}

\bibitem{gamarnikBinaryPerceptron}
Gamarnik D, Kizildag E, Perkins W and Xu C 2021 {\em In preparation\/}

\bibitem{perkins2021frozen}
Perkins W and Xu C 2021 Frozen 1-{RSB} structure of the symmetric {I}sing
  perceptron {\em Proceedings of the 53rd Annual ACM SIGACT Symposium on Theory
  of Computing\/} pp 1579--1588

\bibitem{lesieur2017statistical}
Lesieur T, Miolane L, Lelarge M, Krzakala F and Zdeborov{\'a} L 2017
  Statistical and computational phase transitions in spiked tensor estimation
  {\em 2017 IEEE International Symposium on Information Theory (ISIT)\/} (IEEE)
  pp 511--515

\bibitem{barbier2019optimal}
Barbier J, Krzakala F, Macris N, Miolane L and Zdeborov{\'a} L 2019 {\em
  Proceedings of the National Academy of Sciences\/} {\bf 116} 5451--5460

\bibitem{nishimori2001statistical}
Nishimori H 2001 {\em Statistical physics of spin glasses and information
  processing: an introduction\/} 111 (Clarendon Press)

\bibitem{zdeborova2016statistical}
Zdeborov{\'a} L and Krzakala F 2016 {\em Advances in Physics\/} {\bf 65}
  453--552

\bibitem{thouless1977solution}
Thouless D~J, Anderson P~W and Palmer R~G 1977 {\em Philosophical Magazine\/}
  {\bf 35} 593--601

\bibitem{bolthausen2014iterative}
Bolthausen E 2014 {\em Communications in Mathematical Physics\/} {\bf 325}
  333--366

\bibitem{bayati2011dynamics}
Bayati M and Montanari A 2011 {\em IEEE Transactions on Information Theory\/}
  {\bf 57} 764--785

\bibitem{javanmard2013state}
Javanmard A and Montanari A 2013 {\em Information and Inference: A Journal of
  the IMA\/} {\bf 2} 115--144

\bibitem{bayati2015universality}
Bayati M, Lelarge M and Montanari A 2015 {\em The Annals of Applied
  Probability\/} {\bf 25} 753--822

\bibitem{gerbelot2021graph}
Gerbelot C and Berthier R 2021 {\em arXiv preprint arXiv:2109.11905\/}

\bibitem{decelle2011asymptotic}
Decelle A, Krzakala F, Moore C and Zdeborov{\'a} L 2011 {\em Physical Review
  E\/} {\bf 84} 066106

\bibitem{kesten1966limit}
Kesten H and Stigum B~P 1966 {\em The Annals of Mathematical Statistics\/} {\bf
  37} 1211--1223

\bibitem{mossel2012stochastic}
Mossel E, Neeman J and Sly A 2012 {\em arXiv preprint arXiv:1202.1499\/}

\bibitem{krzakala2012probabilistic}
Krzakala F, M{\'e}zard M, Sausset F, Sun Y and Zdeborov{\'a} L 2012 {\em
  Journal of Statistical Mechanics: Theory and Experiment\/} {\bf 2012} P08009

\bibitem{zdeborova2011quiet}
Zdeborov{\'a} L and Krzakala F 2011 {\em SIAM Journal on Discrete
  Mathematics\/} {\bf 25} 750--770

\bibitem{ricci2019typology}
Ricci-Tersenghi F, Semerjian G and Zdeborov{\'a} L 2019 {\em Physical Review
  E\/} {\bf 99} 042109

\bibitem{semerjian2020recovery}
Semerjian G, Sicuro G and Zdeborov{\'a} L 2020 {\em Physical Review E\/} {\bf
  102} 022304

\bibitem{celentano2020estimation}
Celentano M, Montanari A and Wu Y 2020 The estimation error of general first
  order methods {\em Conference on Learning Theory\/} (PMLR) pp 1078--1141

\bibitem{franz2001ferromagnet}
Franz S, M{\'e}zard M, Ricci-Tersenghi F, Weigt M and Zecchina R 2001 {\em EPL
  (Europhysics Letters)\/} {\bf 55} 465

\bibitem{gamarnik2021inference}
Gamarnik D, K{\i}z{\i}lda{\u{g}} E~C and Zadik I 2021 {\em IEEE Transactions on
  Information Theory\/} {\bf 67} 8109--8139

\bibitem{song2021cryptographic}
Song M~J, Zadik I and Bruna J 2021 {\em Advances in Neural Information
  Processing Systems\/} {\bf 34} 29602--29615

\bibitem{antenucci2019glassy}
Antenucci F, Franz S, Urbani P and Zdeborov{\'a} L 2019 {\em Physical Review
  X\/} {\bf 9} 011020

\bibitem{braunstein2005survey}
Braunstein A, M{\'e}zard M and Zecchina R 2005 {\em Random Structures \&
  Algorithms\/} {\bf 27} 201--226

\bibitem{antenucci2019approximate}
Antenucci F, Krzakala F, Urbani P and Zdeborov{\'a} L 2019 {\em Journal of
  Statistical Mechanics: Theory and Experiment\/} {\bf 2019} 023401

\bibitem{celentano2019fundamental}
Celentano M and Montanari A 2019 {\em arXiv preprint arXiv:1903.10603\/}

\bibitem{sompolinsky1990learning}
Sompolinsky H, Tishby N and Seung H~S 1990 {\em Physical Review Letters\/} {\bf
  65} 1683

\bibitem{chiara2022theory}
Chiara~Angelini M and Ricci-Tersenghi F 2022 {\em arXiv e-prints\/}
  arXiv--2206

\bibitem{mannelli2020marvels}
Mannelli S~S, Biroli G, Cammarota C, Krzakala F, Urbani P and Zdeborov{\'a} L
  2020 {\em Physical Review X\/} {\bf 10} 011057

\bibitem{sarao2020complex}
Sarao~Mannelli S, Biroli G, Cammarota C, Krzakala F, Urbani P and Zdeborov{\'a}
  L 2020 {\em Advances in Neural Information Processing Systems\/} {\bf 33}
  3265--3274

\bibitem{sarao2020optimization}
Sarao~Mannelli S, Vanden-Eijnden E and Zdeborov{\'a} L 2020 {\em Advances in
  Neural Information Processing Systems\/} {\bf 33} 13445--13455

\bibitem{mignacco2021stochasticity}
Mignacco F, Urbani P and Zdeborov{\'a} L 2021 {\em Machine Learning: Science
  and Technology\/} {\bf 2} 035029

\bibitem{barak-steurer}
Barak B and Steurer D 2014 Sum-of-squares proofs and the quest toward optimal
  algorithms {\em Proc. Intl. Congress of Mathematicians (ICM)\/}

\bibitem{krivine}
Krivine J~L 1964 {\em Journal d'Analyse Math{\'e}matique\/} {\bf 12} 307--326

\bibitem{stengle}
Stengle G 1974 {\em Mathematische Annalen\/} {\bf 207} 87--97

\bibitem{Shor-1987-SumOfSquares}
Shor N~Z 1987 {\em Cybernetics\/} {\bf 23} 695--700

\bibitem{Nesterov-2000-SOS}
Nesterov Y 2000 Squared functional systems and optimization problems {\em High
  performance optimization\/} (Springer) pp 405--440

\bibitem{Parrilo-thesis}
Parrilo P~A 2000 Structured semidefinite programs and semialgebraic geometry
  methods in robustness and optimization {Ph.D.} Thesis, California Institute
  of Technology

\bibitem{lasserre}
Lasserre J~B 2001 {\em SIAM J. Optimization\/} {\bf 11} 796--817

\bibitem{ODonnell-2017-SOSNotAutomatizable}
O'Donnell R 2017 {SOS} is not obviously automatizable, even approximately {\em
  8th Innovations in Theoretical Computer Science Conference (ITCS 2017)\/}

\bibitem{RW-2017-BitComplexity}
Raghavendra P and Weitz B 2017 On the bit complexity of sum-of-squares proofs
  {\em 44th International Colloquium on Automata, Languages, and Programming
  (ICALP 2017)\/} vol~80 pp 80:1--80:13

\bibitem{grigoriev2001knapsack}
Grigoriev D 2001 {\em Computational Complexity\/} {\bf 10} 139--154

\bibitem{grigoriev2001linear}
Grigoriev D 2001 {\em Theoretical Computer Science\/} {\bf 259} 613--622

\bibitem{laurent-cut}
Laurent M 2003 {\em Mathematics of Operations Research\/} {\bf 28} 871--883

\bibitem{kunisky-moore}
{Kunisky} D and {Moore} C 2022 {\em arXiv e-prints\/} (\textit{Preprint}
  \eprint{2203.05693})

\bibitem{fawzi-saunderson-parrilo}
Fawzi H, Saunderson J and Parrilo P~A 2016 {\em Mathematical Programming\/}
  {\bf 160} 149--191

\bibitem{cocco-etal}
Cocco S, Dubois O, Mandler J and Monasson R 2003 {\em Phys. Rev. Lett.\/} {\bf
  90}(4) 047205

\bibitem{DL-2009-GeometryCuts}
Deza M~M and Laurent M 2009 {\em Geometry of cuts and metrics\/} (Springer)

\bibitem{Karp-1972-Reducibility}
Karp R~M 1972 Reducibility among combinatorial problems {\em Complexity of
  computer computations\/} (Springer) pp 85--103

\bibitem{parisi79}
Parisi G 1979 {\em Phys. Rev. Lett.\/} {\bf 43}(23) 1754--1756

\bibitem{parisi80}
Parisi G 1980 {\em Journal of Physics A: Mathematical and General\/} {\bf 13}
  L115--L121

\bibitem{laurent-elliptope}
Laurent M and Poljak S 1995 {\em Linear Algebra and its Applications\/} {\bf
  223-224} 439--461

\bibitem{montanari-sen}
Montanari A and Sen S 2016 Semidefinite programs on sparse random graphs and
  their application to community detection {\em Proceedings of the 48th Annual
  Symposium on Theory of Computing\/} p 814–827

\bibitem{goemans-williamson}
Goemans M~X and Williamson D~P 1995 {\em Journal of the ACM\/} {\bf 42}
  1115--1145

\bibitem{khot-etal-unique-games-maxcut}
Khot S, Kindler G, Mossel E and O'Donnell R 2007 {\em SIAM Journal on
  Computing\/} {\bf 37} 319--357

\bibitem{ghosh-etal-sk}
Ghosh M, Jeronimo F, Jones C, Potechin A and Rajendran G 2020 Sum-of-squares
  lower bounds for {S}herrington-{K}irkpatrick via planted affine planes {\em
  Proceedings of 61st Annual Symposium on Foundations of Computer Science
  (FOCS)\/} pp 954--965

\bibitem{isserlis}
Isserlis L 1918 {\em Biometrika\/} {\bf 12} 134--139

\bibitem{kunisky-bandeira-deg4}
Kunisky D and Bandeira A~S 2021 {\em Mathematical Programming\/} {\bf 190}
  721--759

\bibitem{mohanty-raghavendra-xu}
Mohanty S, Raghavendra P and Xu J 2020 Lifting sum-of-squares lower bounds:
  Degree-2 to degree-4 {\em Proceedings of the 52nd Annual Symposium on Theory
  of Computing\/} pp 840–--853

\bibitem{kunisky-deg6}
{Kunisky} D 2020 {\em arXiv e-prints\/} (\textit{Preprint} \eprint{2009.07269})

\bibitem{barak-etal-psuedocalibration}
Barak B, Hopkins S, Kelner J, Kothari P~K, Moitra A and Potechin A 2019 {\em
  SIAM Journal on Computing\/} {\bf 48} 687--735

\bibitem{raghavendra2018high}
Raghavendra P, Schramm T and Steurer D 2018 High dimensional estimation via
  sum-of-squares proofs {\em Proceedings of the International Congress of
  Mathematicians: Rio de Janeiro 2018\/} pp 3389--3423

\bibitem{hopkins2018statistical}
Hopkins S 2018 {\em STATISTICAL INFERENCE AND THE SUM OF SQUARES METHOD\/}
  Ph.D. thesis Cornell University

\bibitem{bandeira-kunisky-wein-pca}
Bandeira A~S, Kunisky D and Wein A~S 2020 Computational hardness of certifying
  bounds on constrained {PCA} problems {\em 11th Innovations in Theoretical
  Computer Science Conference\/} ({\em LIPIcs\/} vol 151) pp 78:1--78:29

\bibitem{krzakala2009hiding}
Krzakala F and Zdeborov{\'a} L 2009 {\em Physical Review Letters\/} {\bf 102}
  238701

\bibitem{bandeira-etal-cuts-colorings}
Bandeira A~S, Banks J, Kunisky D, Moore C and Wein A 2021 Spectral planting and
  the hardness of refuting cuts, colorability, and communities in random graphs
  {\em Proceedings of the 34th Conference on Learning Theory\/} pp 410--473

\bibitem{banks-kleinberg-moore}
Banks J, Kleinberg R and Moore C 2019 {\em SIAM Journal on Computing\/} {\bf
  48} 1098--1119

\bibitem{barak-kelner-steurer-rounding}
Barak B, Kelner J~A and Steurer D 2014 Rounding sum-of-squares relaxations {\em
  Proceedings of the Forty-Sixth Annual ACM Symposium on Theory of Computing\/}
  STOC '14 p 31–40

\bibitem{barak-moitra-matrix-completion}
Barak B and Moitra A 2016 Noisy tensor completion via the sum-of-squares
  hierarchy {\em 29th Annual Conference on Learning Theory\/} pp 417--445

\bibitem{hopkins-shi-steurer-tensor-pca}
Hopkins S~B, Shi J and Steurer D 2015 Tensor principal component analysis via
  sum-of-square proofs {\em Proceedings of the 28th Conference on Learning
  Theory\/} pp 956--1006

\bibitem{richard2014statistical}
Richard E and Montanari A 2014 A statistical model for tensor pca {\em Advances
  in Neural Information Processing Systems\/} pp 2897--2905

\bibitem{hopkins-etal-fast-spectral}
Hopkins S~B, Schramm T, Shi J and Steurer D 2016 Fast spectral algorithms from
  sum-of-squares proofs: Tensor decomposition and planted sparse vectors {\em
  Proceedings of the Forty-Eighth Annual ACM Symposium on Theory of
  Computing\/} p 178–191

\bibitem{hopkins2017power}
Hopkins S~B, Kothari P~K, Potechin A, Raghavendra P, Schramm T and Steurer D
  2017 The power of sum-of-squares for detecting hidden structures {\em 2017
  IEEE 58th Annual Symposium on Foundations of Computer Science (FOCS)\/}
  (IEEE) pp 720--731

\bibitem{anandkumer-tensor-pca}
Anandkumar A, Ge R and Janzamin M 2017 {\em J. Machine Learning Research\/}
  {\bf 18} 752–791

\bibitem{wein-alaoui-moore}
Wein A~S, Alaoui A~E~K and Moore C 2019 {\em Proceedings of 60th Annual
  Symposium on Foundations of Computer Science\/}  1446--1468

\bibitem{YedidiaFreemanWeiss}
Yedidia J, Freeman W and Weiss Y 2001 {\em Mitsubishi Elect. Res. Lab.\/}

\bibitem{kikuchi}
Kikuchi R 1951 {\em Physical Review\/} {\bf 81}(6) 988--1003

\bibitem{yedidia-kikuchi-nips}
Yedidia J~S, Freeman W and Weiss Y 2000 Generalized belief propagation {\em
  Advances in Neural Information Processing Systems\/}

\bibitem{YedidiaFreemanWeiss-kikuchi}
Yedidia J, Freeman W and Weiss Y 2001 {\em Mitsubishi Elect. Res. Lab.\/}

\bibitem{spectral-redemption}
Krzakala F, Moore C, Mossel E, Neeman J, Sly A, Zdeborov{\'a} L and Zhang P
  2013 {\em Proceedings of the National Academy of Sciences\/} {\bf 110}
  20935--20940

\bibitem{bordenave-lelarge-massoulie}
Bordenave C, Lelarge M and Massoulie L 2015 Non-backtracking spectrum of random
  graphs: Community detection and non-regular {R}amanujan graphs {\em
  Proceedings 56th Annual Symposium on Foundations of Computer Science\/} pp
  1347--1357

\bibitem{saade-bethe-hessian}
Saade A, Krzakala F and Zdeborov\'{a} L 2014 Spectral clustering of graphs with
  the bethe hessian {\em Advances in Neural Information Processing Systems\/}
  pp 406--414

\bibitem{bhattiprolu_et_al}
Bhattiprolu V, Guruswami V and Lee E 2017 Sum-of-squares certificates for
  maxima of random tensors on the sphere {\em Proceedings of APPROX/RANDOM
  2017\/} pp 31:1--31:20

\bibitem{raghavendra-rao-schramm}
Raghavendra P, Rao S and Schramm T 2017 Strongly refuting random {CSP}s below
  the spectral threshold {\em Proceedings of the 49th Annual Symposium on
  Theory of Computing\/} p 121–131

\bibitem{guruswami-kothari-manohar}
Guruswami V, Kothari P~K and Manohar P 2022 Algorithms and certificates for
  boolean {CSP} refutation: Smoothed is no harder than random {\em Proceedings
  of the 54th Annual Symposium on Theory of Computing\/} p 678–689

\bibitem{bandeira-etal-franz-parisi}
{Bandeira} A~S, {El Alaoui} A, {Hopkins} S~B, {Schramm} T, {Wein} A~S and
  {Zadik} I 2022 {\em arXiv preprint\/} (\textit{Preprint} \eprint{2205.09727})

\end{thebibliography}

\end{document}